# Chirped temporal solitons in driven optical resonators

**Christopher Spiess, Qian Yang, Xue Dong, Victor G. Bucklew, and William H. Renninger**
*Institute of Optics, University of Rochester, Rochester, NY 14627, USA*
william.renninger@rochester.edu

**Abstract**
Temporal solitons in driven microresonator, fiber-resonator, and bulk enhancement cavities enable attractive optical sources for spectroscopy, communications, and metrology. Here we present theoretical and experimental observations of a new class of temporal optical soliton characterized by pulses with large and positive chirp in normal dispersion resonators with strong spectral filtering. Numerical simulations reveal stable waveforms over a wide new range of parameters including highly chirped pulses at large drive powers. Chirped temporal solitons matching predictions are observed in experiments with normal dispersion fiber resonators strongly driven with nanosecond pulses. Scaling laws are developed and provide simple design guidelines for generating chirped temporal solitons in bulk- and micro-resonator, in addition to fiber-resonator platforms. The relationship between the chirped solutions and other stable waveforms in normal and anomalous dispersion resonators is examined. Chirped temporal solitons represent a promising new resource for frequency-comb and ultrashort-pulse generation.

**Introduction**
Frequency combs can be generated in optical resonators driven by a continuous-wave laser. In these systems, bandwidth is generated through Kerr-mediated parametric four-wave mixing and stability is achieved through nonlinear self-organizing processes. In driven fiber cavities, early studies of modulation instability demonstrated evidence of pattern formation [1–3], and more recent studies have focused on long-range interactions [4], spatiotemporal instabilities [5], temporal tweezing [6], and applications such as all-optical buffering [7,8]. In parallel, researchers have established micron-scale resonators as compact, simple, and low-power sources of frequency combs with large frequency spacings. Microresonator source development has attracted considerable interest for applications in waveform synthesis, high-capacity telecommunications, astrophysical spectrometer calibration, atomic clocks, and dual-comb spectroscopy [9,10]. Microresonator frequency combs have been demonstrated in whispering gallery [11,12] and on-chip [13,14] cavities, with high performance combs spanning an octave or more [15–17]. Low noise and broadband coherence are critical for frequency-comb applications and require specific consideration for driven-resonator sources [18,19]. Specifically, the phases between the cavity modes must have a well-defined relationship- i.e. the driven cavity must be mode-locked.

In laser cavities, with an active gain medium, mode-locking occurs through the formation of optical solitons, which are pulses that self-stabilize in the presence of Kerr optical nonlinearity and anomalous group-delay dispersion (GDD) [20]. Optical solitons are an attractive mechanism for stabilizing broadband frequency combs in driven passive cavities as well. Solitons in these systems also ensure that broadband cavity losses are counter-balanced by the single-frequency drive source. Driven-cavity solitons were first observed in fiber resonators [7], shortly thereafter in microresonators [21], and most recently in bulk enhancement cavities [22]. The close relationship between driven-cavity soliton mode-locking and laser-cavity soliton mode-locking [23] suggests new potential mechanisms for stable pulse and frequency-comb generation in driven-cavity systems. For example, in mode-locked lasers featuring normal dispersion and a spectral filter, a distinct class of highly-chirped soliton can be generated [24–27]. In addition to the scientific importance of novel highly dissipative soliton formation, the chirped-pulse laser soliton has benefited applications by extending pulse generation to normal dispersion systems, enabling large pulse energies [28], and simplifying amplifier setups [29]. The impact that chirped solitons have had for mode-locked lasers motivates the search for analogous solutions in driven-resonator systems. While several nonlinear solutions have been analyzed in driven-cavities, including Turing patterns [30], breathing pulses [30], and soliton crystals [31] in anomalous dispersion cavities and dark solitons [32–35], bright solitons [36], platicons [35,37,38], and switching waves [32,39] in normal dispersion cavities [40], an analogous chirped temporal soliton has not been observed.

Here we observe chirped temporal solitons in driven resonators theoretically and experimentally in normal dispersion fiber cavities with a spectral filter. A driven normal-dispersion resonator with effective spectral filtering has been examined previously [41], but the filter was not specifically designed and chirped pulses were not observed. Numerically simulated resonators with suitable spectral filtering are found to support stable pulses with chirp that corresponds to more than twice the linear dispersion of the cavity, which indicates that the spectral phase is the result of nonlinear pulse formation. Chirped solitons, in agreement with predications, are observed experimentally in long normal dispersion fiber cavities driven with nanosecond pulses. General scaling laws are developed for chirped temporal solitons in driven resonator systems. Chirped temporal solitons enable a broad new range of system and performance parameters that complement currently available techniques for ultrashort pulse and frequency-comb generation.

**Theory**
A passive fiber resonator is designed to support chirped temporal solitons. The cavity is designed based on an analogous chirped-pulse mode-locked laser in which the laser amplifier is replaced with a continuous-wave drive source. The passive cavity consists of normal dispersion fiber, losses, a drive source, and a spectral filter (Fig. 1a). Numerical simulations are developed to determine if this design can support chirped temporal solitons (see Methods). The spectral filter (a Gaussian profile with 4-nm bandwidth) is chosen based on the requirements for a mode-locked fiber laser with the same cavity length (52.5 m in this case) [26,42]. After fixing the filter, cavity length, and losses, the simulations are examined as a function of the remaining variables: the incident drive power and frequency (cavity detuning). First, the trivial continuous-wave solutions and solutions that do not converge are indicated by white regions in Fig. 1b. To identify non-trivial solutions, multiple characteristics are examined, including the spectral bandwidth, peak power, chirp, and number of prominent intensity peaks (see Supplementary Information, Section 3). The number of peaks in the converged waveform provides particularly good contrast between different solution types (Fig. 1b). A variety of stable non-trivial solutions are observed, including chirped pulses, dark pulses, switching-waves, and Turing patterns. In Fig. 1b, the Turing patterns, with more than eight intensity peaks, are indicated with red points, and the dark pulses, switching waves, and chirped pulses, which can occur with a single peak, are indicated with blue points. The different nonlinear solutions can also coexist (see Supplementary Information, Section 4). The chirped temporal solitons of interest exist over a broad range of detuning and drive values, including both signs of detuning. The minimum (threshold) intracavity drive power for which chirped pulses are observed is 5 W. The dynamics and stability regions of all of the solutions change significantly when the spectral filter bandwidth changes. For example, for chirped pulses, the threshold drive power decreases with narrower filter bandwidths (see Supplementary Information, Section 5). Moreover, the chirped pulses are not observed at all with broadband or without spectral filtering (see Supplementary Information, Section 5).

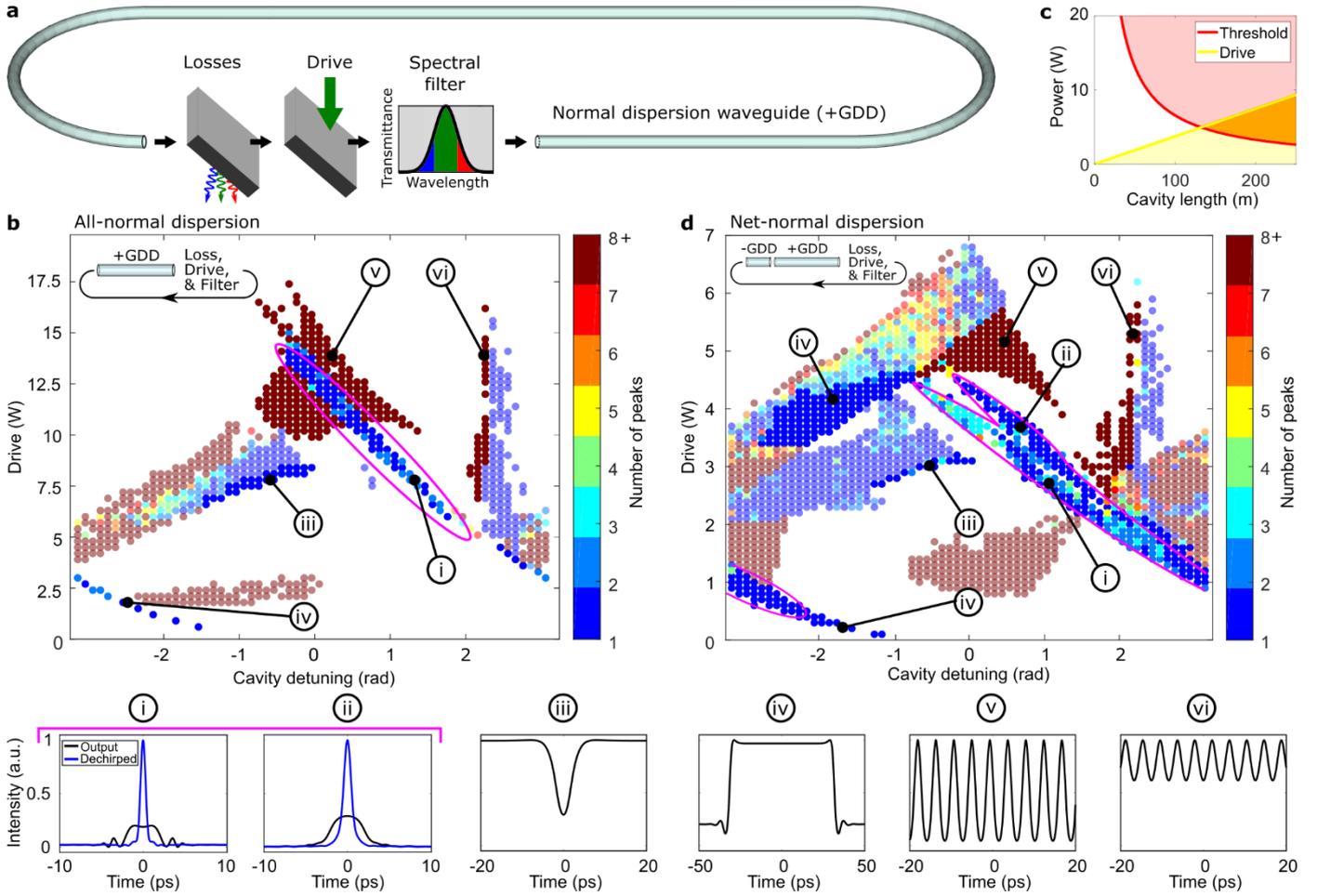

Fig. 1: **Cavity design for chirped temporal soliton generation and numerical results. a** Illustration of the key components for observing chirped pulses in driven resonators, including a normal dispersion waveguide, losses, drive, and a spectral filter. **b** Converged solutions as a function of drive power and detuning for an all-normal dispersion cavity. The color map represents the number of prominent peaks on the stable waveform. Specific solutions are indicated with Roman numerals. i, ii Chirped pulse are observed in a well-defined region of parameter space (pink outline), and iii dark pulses, iv switching waves, and v, vi Turing waves are observed elsewhere. **c** The minimum cavity length required for observing chirped solitons (orange) is determined by the decreasing threshold power (red) and the increasing peak drive power (yellow) as a function of length. **d** Converged solutions as a function of drive power and detuning comparable to (b) for the cavity with large net-normal dispersion developed experimentally. See Methods for cavity parameters.

The chirped pulse solitons rapidly converge to a steady-state in the cavity (Fig. 2a and Supplementary Information, Section 1). In the example illustrated in Fig. 2, the picosecond pulses exhibit a positive chirp corresponding to 1 $ps^2$ of group-delay dispersion (Fig. 2b and Methods). This corresponds to more than double the group-delay dispersion of the normal dispersion fiber in the cavity, which indicates that the chirp is the result of nonlinear pulse formation. The dechirped pulse peak power is enhanced by a factor of ~4 from its value directly out of the cavity (Fig. 2c). The pulses can be compressed to 0.9 ps, which is close to the transform-limited pulse duration of 0.82 ps and indicates that the chirp is nearly linear. The duration and chirp of chirped temporal solitons vary depending on the bandwidth of the spectral filter (see Supplementary Information, Section 6).

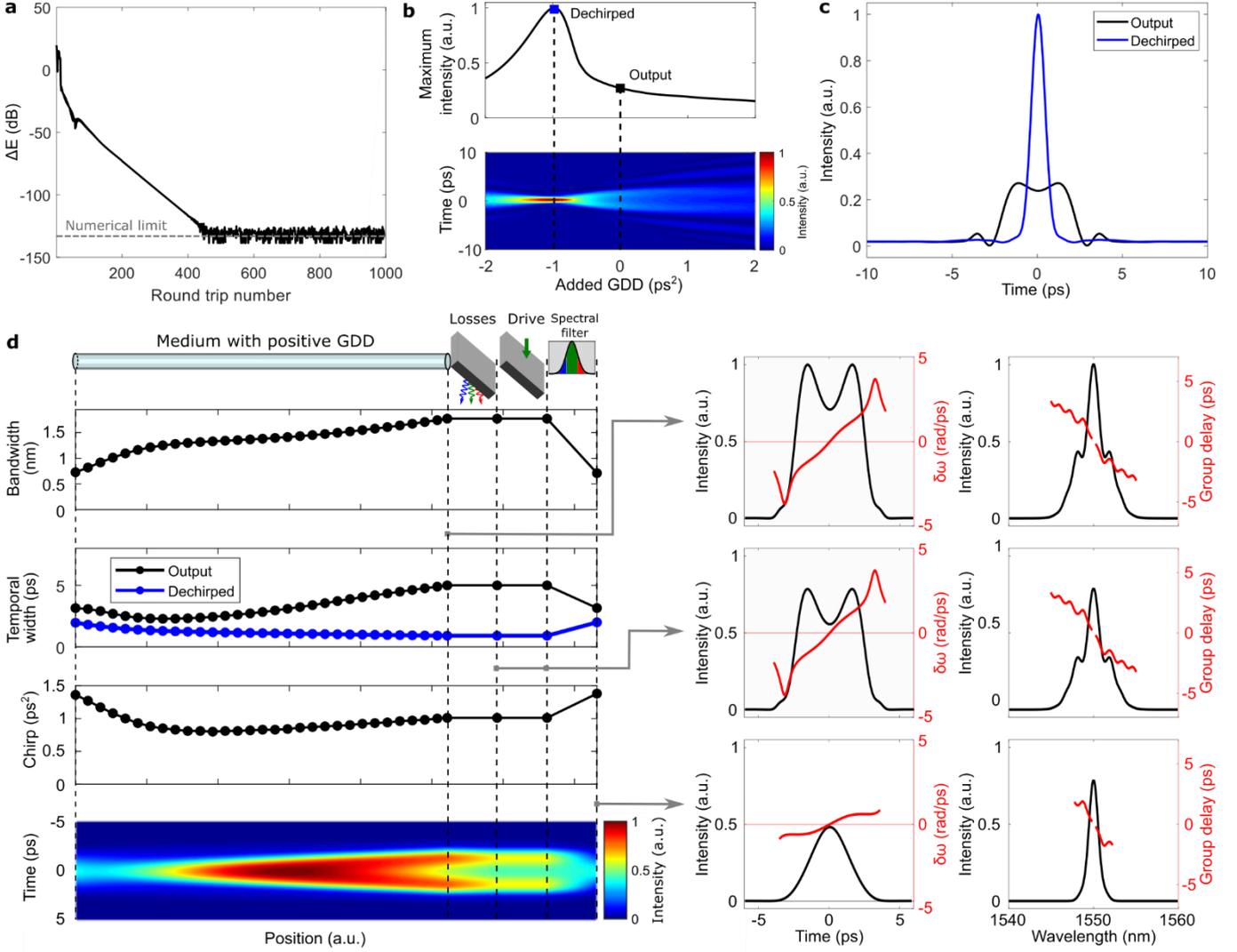

Fig. 2. **Characteristics of chirped temporal solitons in driven optical cavities. a** Numerical convergence of the pulse energy difference between subsequent round trips, Δ$E$, to a stable numerically-limited steady-state value. **b** The change in the pulse and peak intensity as a function of group-delay dispersion applied after the cavity, indicating continuous compression with anomalous dispersion with a maximum at GDD = -1 ps$^2$. **c** The chirped cavity output (black) and dechirped (blue) pulses from b. **d** Evolution of steady-state chirped temporal soliton bandwidth (full width at half the maximum, FWHM), temporal width (FWHM), chirp (defined by the GDD required to maximize the pulse intensity, with the opposite sign), and pulse intensity in the cavity. The FWHM of the pulse after dechirping the pulse at each position of the cavity is plotted in blue. The associated pulse intensity, instantaneous frequency, power spectrum, and group delay from the indicated locations in the cavity are plotted on the right. The positive slope of the instantaneous frequency, $\delta\omega$, and the negative slope of the spectral group delay correspond to the soliton chirp.

The spectral bandwidth, temporal duration, and chirp magnitude evolve nonlinearly in the cavity (Fig. 2d). In the normal dispersion waveguide, the spectrum experiences a net broadening due to Kerr self-phase modulation. The temporal width increases primarily owing to dispersive propagation in the normal dispersion waveguide. The spectral filter reduces the spectral bandwidth and the pulse duration is also reduced when the high and low frequencies in the leading and trailing edges of the pulse are attenuated. The pulse is highly chirped at every point in the cavity. This can also be seen from the positive slope of the instantaneous frequency on the right of Fig. 2. The dissipative drive and losses have a negligible effect on the pulse, spectrum, and chirp. Overall, the qualitative balance of physical effects is similar to that in chirped-pulse mode-locked lasers, in which pulse and spectral broadening are also counteracted by spectral filtering [26,42]. However, we note that the continuous-wave background (absent in mode-locked lasers) has a time dependent phase relationship with the chirped pulse, which results in an oscillatory structure in the time domain that can complicate the interpretation of the evolution.

Simple scaling laws can be developed to identify the cavity parameters necessary to obtain chirped pulse solutions in driven cavity systems. This is achieved starting with the well-established mean-field model for the driven-cavity system, the Lugiato-Lefever equation (LLE). To account for additional spectral filtering a term that represents a distributed Gaussian spectral filter is added (see Supplementary Information, Section 7). The normalized equation can be defined by the following three unitless coefficients related to the drive power, spectral filter bandwidth, and drive detuning:

$$D_0 = D\frac{\gamma L}{\alpha^3}, \quad f_0 = f\sqrt{L|\overline{\beta_2}|}, \quad \text{and} \quad \delta_0 = \frac{\delta}{\alpha}, \tag{1}$$

where $D$ is the intracavity drive, $\delta$ the frequency detuning, $f$ is the filter bandwidth, L is the cavity length, $\alpha$ accounts for the cavity losses, $\gamma$ is the nonlinear coefficient, and $\overline{\beta_2}$ is the average group-velocity dispersion (GDD divided by L). If chirped temporal solitons are known to be stable in a particular cavity with specific values of $D_0$, $f_0$, and $\delta_0$, stable chirped solitons (with peak power and duration scaled appropriately) can also be obtained for a different cavity as long as the values for the unitless coefficients do not change. While chirped pulses may be stable for many different values of these parameters, only one set is needed for design. The first relationship from Eq. 1 reveals that the required drive power has a cubic dependence on the total cavity loss and an inverse linear dependence on the total cavity nonlinearity (see Supplementary Information, Section 8). The second relationship conveys that the filter bandwidth must scale inversely with the square root of the total group delay dispersion (see Supplementary Information, Section 5). The third relationship suggests that the same solution can be recovered if the relative drive frequency scales linearly with the cavity loss. These simple relationships provide powerful design guidelines for obtaining chirped pulse solitons in normal-dispersion driven cavities with a filter.

The chirped temporal solitons observed numerically require drive powers that are challenging to obtain experimentally (Fig. 1b). Numerically, while stable solutions can be obtained with low drive powers, they exist over a narrow range of parameters and may be challenging to observe in practice. In contrast, at higher drive powers, chirped pulses are stable over a large range of detuning values and may be more readily observed. This is the case, for example, in the all-normal dispersion cavity simulated in Fig. 1b, with 10-W of intra-cavity drive power. However, since we are experimentally limited to 0.1 W of intra-cavity drive power, an advance is needed. Therefore, to experimentally observe chirped solitons we design the cavity to have a reduced drive threshold and achieve higher drive powers with a pulsed drive source. Driving passive cavities with pulses is a recently established technique for achieving higher peak drive powers[5,43,44]. For this technique, the continuous-wave drive is modulated into a nanosecond pulse with the repetition rate of the cavity, and then amplified. This enhances the drive power by an amount that corresponds to the duty cycle of the drive pulse train (drive pulse duration divided by the cavity round trip time). For a fixed minimum drive pulse duration, this results in linear enhancement in drive power with cavity length (see Supplementary Information, Section 9). In addition, Eq. 1 reveals that the required drive power can be reduced linearly with an increase in the total cavity length. Considering both the reduced threshold and the increased drive, we find that stable chirped pulses should be observable for cavity lengths longer than 150 m, which corresponds to a three-fold increase in drive power and a three-fold decrease in drive threshold (Fig. 1c). However, an increased cavity length would also increase the total dispersion, with a corresponding reduction in the bandwidth of the output pulse (see Supplementary Information, Section 7). To avoid this loss of bandwidth, the length of the cavity is increased without changing the total dispersion with a dispersion-map consisting of two fibers with opposite signs of dispersion (dispersion management). A dispersion-managed approach allows for independent control of the drive threshold in driven fiber resonators.

A suitable dispersion-managed fiber cavity is numerically modeled to confirm that chirped temporal solitons are stable in an experimentally compatible system. To more accurately represent experimental parameters, the exact super-Gaussian profile of the bandpass spectral filter and the third-order dispersion of the fibers are incorporated in the model. Simulations are run for a 150-m cavity with the same total dispersion as in the all-normal dispersion cavity from Fig. 1b. With only subtle differences from the all-normal dispersion system (see details in Supplementary Information, Section 6), the drive threshold scales as expected, and stable chirped temporal solitons are observed with experimentally accessible drive powers of ~3 W (Fig. 1d). The numerical results validate the dispersion-managed cavity approach and motivate experimental investigation.

**Experiment**
Following the results of numerical simulations, a fiber resonator is designed to support chirped temporal solitons (see Fig. 3a and Methods). The cavity consists of 150-m total length of single-mode fiber with large net-normal dispersion and a 4.25-nm fiber-format spectral filter. The drive is pulsed to enable access to high intra-cavity powers. The cavity resonance deviates significantly from a Lorentzian profile at average drive powers larger than 0.3 W, indicating the nonlinear nature of the resonance (Fig. 3d). The equivalent numerical resonance agrees well with the experiment (Fig. 3h). See Methods and Supplementary Information, Section 10 for additional information.

Stable and reproducible chirped-pulse solutions are observed with appropriate adjustment of the drive frequency, power, polarization, and pulse period. The output optical spectrum features a unique profile characteristic of the simulated chirped pulses from Region ii in Fig. 1d (Fig. 3c). The spectra quantitatively agree with theory including small sidebands 2-nm shifted from the center wavelength and an rms-bandwidth of 1.2 nm (Fig. 3g). Small variations of these values can also be observed (see Supplementary Information, Section 11). The spectrum is measured after a fiber Bragg filter, resulting in the modulation in the center of the spectrum. The pulse train, observed with an oscilloscope, consists of pulses regularly spaced in time with the cavity round-trip period and energy fluctuations of less than 1% (Fig. 3b). To evaluate the output pulse chirp, the pulses are amplified and dechirped by a grating-pair compressor and measured with a collinear intensity autocorrelator (see Supplementary Information, Section 12). The autocorrelation width reduces to a minimum value before increasing again with further application of anomalous dispersion, which indicates positive chirp (Fig. 3e,f). The minimum duration (1.08 ps FWHM) corresponds to a group-delay dispersion of 1.5 ps², which is three times the cavity group-delay dispersion and indicates that the chirp is the result of nonlinear pulse formation. Comparable autocorrelations of numerically simulated chirped temporal solitons agree well with the experimental observations (Fig. 3i,j).

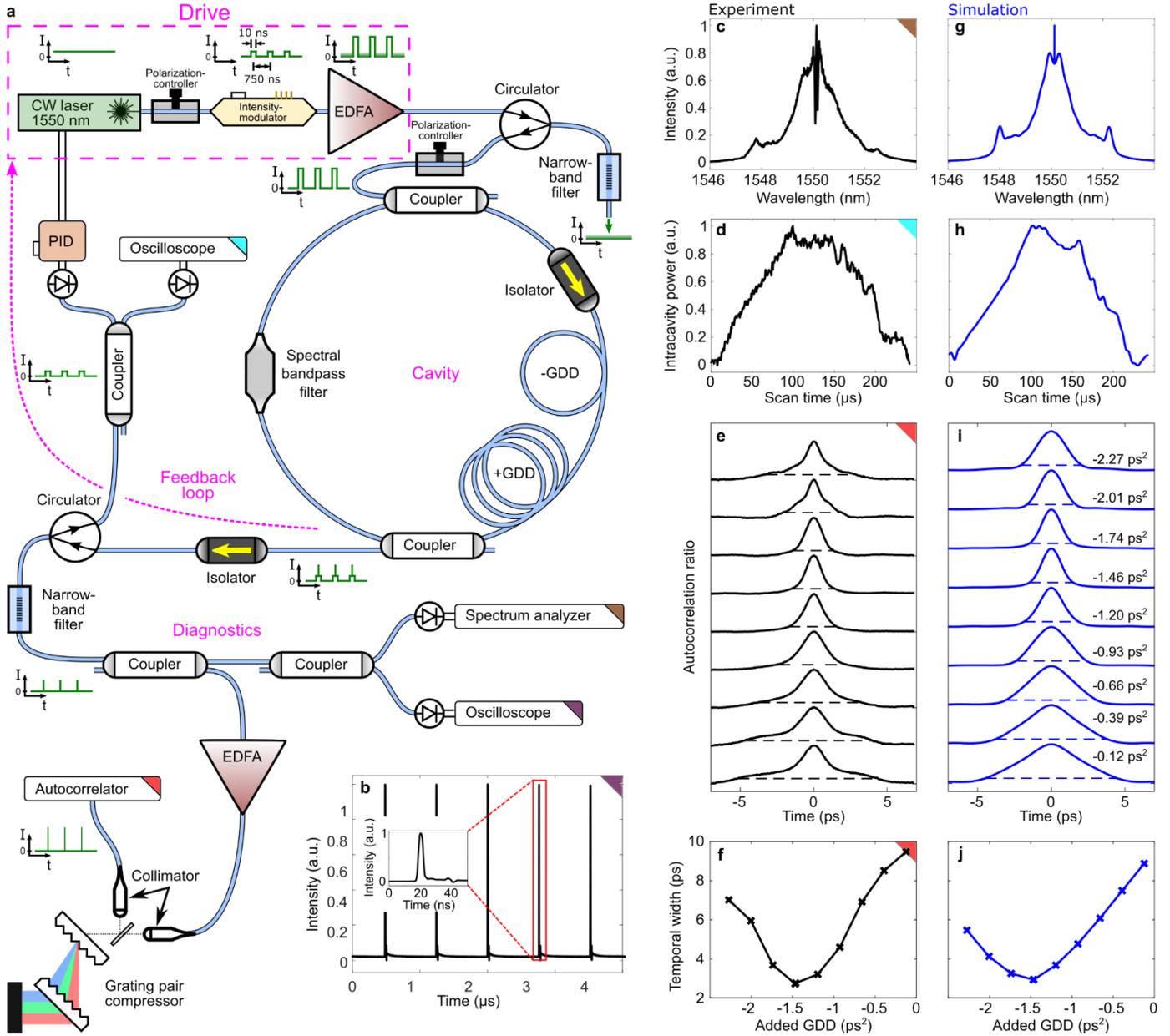

Fig. 3. **Experimental schematic and observations of chirped temporal solitons**. **a** Experimental schematic depicting the modulated and amplified drive, the fiber cavity, the drive feedback loop, and diagnostics. The temporal waveform intensity is indicated throughout in green. Experimentally observed **b** pulse train, **c** spectrum, **d** cavity resonance at 1.8-W average power, and **e-f** autocorrelation measurements. The resonance sweep is measured after a narrow passband filter and the spectrum, pulse train, and autocorrelations are measured after a corresponding narrow notch filter. Autocorrelations are measured after amplification as a function of the GDD from a grating pair compressor with the temporal pulse-width (at 15% of the maximum) plotted in **f**. A colored triangle in the upper right corner of the experimental data indicates the location and type of equipment used to obtain it with a matching triangle in the schematic. Comparable results from numerical simulations are plotted for the **g** spectrum, **h** resonance sweep, and **i-j** autocorrelation measurements, in blue.

**Discussion**

The presented theoretical model does not account for the role of polarization dynamics in the cavity. Interactions between distinct polarizations can lead to modulation instability, which will affect the initiation of pulses in the cavity. Polarization dynamics can also lead to intensity dependent losses and are likely to influence pulse formation. Experiments suggest the relevance of polarization dynamics because pulse initiation and the character of the steady-state solutions are affected by the orientation of the polarization controllers. It will be valuable to account for the effects of polarization with a vectorized numerical model which includes the orthogonal fiber polarization states as well as linear and nonlinear coupling [45].

Chirped temporal solitons in passive normal dispersion resonators with a spectral filter are related to previously investigated nonlinear solutions in resonators without a filter (see Supplementary Information, Section 13). At low powers in the fiber resonator and the LLE, dark pulses with a well-defined single trough are stable [30,32]. At higher drive powers and detuning values, the width and complexity of the dark pulse increases. Complex dark pulses with a long duration appear less like dark pulses and more like interlocking switching waves, which has been established as the theoretical basis for this class of solutions in the LLE [32]. At higher powers, dark pulses become unstable in the LLE and in simulations of the fiber cavity. However, when a weak spectral filter is added to the resonator (a 20-nm bandwidth for the 52.5-m cavity considered here), the long complex dark pulses (switching waves) begin to shift in parameter space and are stable at an order of magnitude higher powers. With a stronger 4-nm filter, the switching waves shift to even higher drive powers. With this strong filter, chirped pulses begin to become stable along the same line in parameter space, but at drive powers that are as much as a hundred times higher than the powers needed for dark pulses in resonators without a filter. These bright pulses can no longer be interpreted as long-duration dark pulses because they have a well-defined pulse width which is maintained regardless of the number of pulses observed. The bandwidth of the chirped pulse is broad compared to the other solutions and the spectral phase has a clear quadradic component, which defines the chirp. However, while these solutions are distinguished from dark pulses, the interlocking switching-wave description may be relevant and merits further investigation. Interestingly, chirped pulse solitons in mode-locked lasers can also be interpreted as the intersection of switching-wave solutions. The cubic-quintic Ginzburg Landau equation that governs mode-locked lasers possesses chirped pulse solutions that are well described analytically by the intersection of two propagating front solutions [46]. This common description may relate chirped temporal solitons in passive resonators to chirped solitons in mode-locked lasers further, in addition to their comparable chirp, parameter requirements, and evolution. Further research into the formation and stability mechanisms of chirped temporal solitons in passive cavities and their relationship to other normal dispersion solutions is needed.

Chirped temporal solitons in normal dispersion resonators with a spectral filter have a higher drive threshold power than traditional solitons. In this work, chirped pulses were observed with 2-W of average power (before the cavity), through the combination of a long fiber cavity and a pulsed drive. However, certain applications may require smaller average powers for pulse (comb) generation. The peak power of the drive is determined by the average power times the ratio of the drive pulse duration over the cavity round trip time. This peak power is currently limited by the duration of the drive pulse, which is limited by our pulse generator to >10-ns. This duration can be decreased to 100-ps with a suitable generator, which will decrease the average power threshold of the system by two orders of magnitude. In addition, the threshold can be decreased further with narrower bandwidth spectral filters (see Supplementary Information, Section 5). With both improvements, chirped temporal solitons can be generated in fiber resonators with mW drive power.

Broad bandwidth is important for frequency-comb as well as for ultrashort pulse applications. The chirped pulses observed here have bandwidth corresponding to picosecond pulse durations. The soliton bandwidth can be increased by decreasing the total dispersion and by applying a correspondingly larger bandwidth spectral filter (see Supplementary Information, Section 5). The scaling laws predict that the soliton bandwidth increases in proportion to the inverse of the square root of the cavity group delay dispersion if the spectral filter bandwidth is increased with the same proportion. In other words, ten times broader soliton bandwidth should be possible with a cavity group delay dispersion that is one hundred times smaller and a spectral filter with ten times broader bandwidth than the present configuration.

For a given bandwidth, the energy of the pulse determines important parameters for applications, including the pulse peak power, the frequency comb power-per-comb line, and the conversion efficiency. Since chirped solitons in mode-locked lasers have higher energies than solitons in anomalous dispersion laser cavities, it will be important to determine if a similar benefit can be achieved for passive cavities. In passive cavities the pulse energy is challenging to measure accurately because it is difficult to accurately determine the total number of pulses and residual continuous-wave background complicates the interpretation of average power measurements. These challenges can potentially be addressed through seeding the cavity with an external source and with background management techniques; this is the subject of on-going research. In addition, numerical simulations can provide important information about the pulse energy, including the potential enhancements compared to traditional solitons and opportunities for further increases (see Supplementary Information, Section 14). For example, the energy of the simulated chirped solitons corresponding to experimental observations is 25 pJ. In a controlled numerical comparison between traditional solitons in anomalous dispersion cavities and chirped solitons in normal dispersion cavities we find that chirped pulses can have at least seven times more energy (see Supplementary Information, Fig. S14). By driving normal dispersion resonators with higher powers, the energy of stable chirped temporal solitons can be increased by at least two times more. Alternatively, optimizing for peak power instead of energy, we find that chirped-temporal solitons can support at least an order of magnitude higher peak powers than traditional solitons given an equivalent magnitude of group-delay dispersion. The simulated results are encouraging, highlight the promise of chirped temporal solitons for applications, and motivate further research.

In passive resonators, the resonance frequencies are sensitive to environmental perturbations including vibrations and temperature. Moreover, the drive laser frequency must be locked with respect to these resonances. Therefore, the stability of frequency-comb generation is proportional to the strength of environmental perturbations and the quality of the frequency locking mechanism. The resonator investigated in this study features minimal temperature and vibration control, a limited laser frequency tuning range, a free-running drive repetition rate, and a single-stage side-lock PID feedback loop. Stable frequency-comb generation in this nonideal configuration lasts for several minutes. However, with several improvements, including temperature and vibration control, an additional feedback loop to control for large frequency changes by thermally tuning the laser, locking the drive repetition rate to the cavity, and peak-locking techniques, stable frequency combs should be generated over significantly longer periods, with minimal variation.

Chirped temporal solitons represent a new class of stable nonlinear waveforms in driven resonator systems. The present study focuses specifically on fiber resonators, but the results are general and can be applied to any passive resonator platform using the scaling laws given by Eq. 1. Passive resonators enable femtosecond pulse generation at wavelengths not accessible by traditional mode-locked lasers and may complement these systems. The chirped temporal soliton extends ultrashort pulse generation to normal dispersion systems, enables new performance regimes, and represents a valuable new solution for frequency comb and ultrashort pulse generation and associated applications.

## Methods

**Numerical simulations.** Numerical simulations are developed to determine if a cavity consisting of normal dispersion fiber, losses, a drive source, and a spectral filter can support chirped temporal solitons. The fiber is modeled by a detuned nonlinear Schrödinger equation incorporating dispersive and nonlinear phase modulations in addition to a term corresponding to the frequency detuning of the drive from the peak of the cavity resonance [47]. The fiber section is simulated with the standard split-step Fourier technique with the dispersive effects calculated in the Fourier domain and the nonlinear effects solved with a 4th order Runge-Kutta method. After the fiber section, the loss, drive, and spectral filter are added as lumped elements. The initial cavity under consideration consists of 52.5 m of fiber with $n_2 = 3.2 \times 10^{-20} \frac{m^2}{W}$, $\beta_2 = 9688 \frac{fs^2}{m}$, and mode-field diameter of d=8.1 µm, total losses of 1.05 dB (as predicted for a typical experimental cavity), and a Gaussian spectral filter with 4-nm spectral bandwidth.

Solutions are identified as stable if the field converges to a steady-state after a finite number of iterations around the cavity (see Supplementary Information, Section 1). For example, the chirped pulse solutions from Fig. 2 are stable, and the mean intensity of the electric field converges to a constant numerically limited value after 500 round-trips, which corresponds to 125 µs for the 50-m cavity with a 4-MHz repetition rate (Fig. 2a). The cavity is seeded by Gaussian or random electric field initial conditions and multiple round trips are simulated. The combination of 9 noisy and 6 Gaussian initial fields results in a sufficient number of stable solutions to establish clear boundaries between different solution types (see Supplementary Information, Section 2).

The chirp is evaluated through the application of anomalous GDD to the pulse, in keeping with the experimental practice of 'dechirping' the pulse with a grating pair dispersive compressor. The chirp magnitude in units of ps$^2$ is determined by the GDD required to maximize the pulse peak intensity (Fig. 2b). In the example from Fig. 2, with a cavity with a 2-nm spectral filter (drive 11.4 W and detuning 1.36 rad), the intensity is smoothly maximized and indicates a positive chirp that corresponds to a GDD of 1 ps$^2$.

For the dispersion-managed simulations the normal dispersion fiber is modeled as above, the anomalous dispersion fiber is modeled with $n_2 = 3.2 \times 10^{-20} \frac{m^2}{W}$, $\beta_2 = -22946 \frac{fs^2}{m}$, and mode-field diameter of d=10.4 µm, and the third-order dispersion for both fibers is given by $\beta_3 = 0.1 \times 10^6 \frac{fs^3}{m}$. The spectral filter has a 12th order super-Gaussian response with a full-width at half-maximum bandwidth of 4.25 nm.

To numerically model the resonance as a function of frequency for comparison with experiments we use a noise-seeded simulation in which the detuning is varied after each round trip at a rate determined by the experimental sweep time. The continuous-wave intensity is averaged over 10 different random-intensity initial fields and plotted at each value of detuning (Fig. 3h). See Supplementary Information, Section 10 for additional information.

**Experimental setup and parameters.** Experimentally, following the results of numerical simulations, a fiber resonator is designed to support chirped temporal solitons. The cavity consists of a total length of 150-m single-mode fiber with a net-dispersion that corresponds to 52.5-m of normal dispersion fiber (with $\beta_2 = 9688 \frac{fs^2}{m}$). An isolator ensures unidirectional operation and suppresses Brillouin scattering. The drive is coupled into the cavity with a 5% fiber-format coupler, and the output is coupled out from a distinct 2% fiber coupler. A 4.25-nm 12th order super-Gaussian fiber-format spectral filter is spliced into the cavity after the output coupler. The drive consists of an intensity-modulated narrow-line fiber laser. The intensity-modulator is driven with 10-ns pulses with a 750-ns period matching that of the fiber cavity. The modulated drive is amplified, and residual amplified spontaneous emission is filtered out with a 20-GHz fiber-Bragg notch filter. 2 W of average power is available before the input fiber coupler. Polarization controllers are used to control the polarization state separately before the intensity modulator and before the fiber cavity. The drive frequency is locked to the cavity resonance with a PID control circuit using the output continuous-wave power as an error signal. The PID circuit enables control of the frequency offset, or detuning, from the cavity resonance. To measure the cavity resonance, the continuous-wave output power is measured as the laser frequency is swept through the cavity resonance. The laser frequency is periodically swept through the cavity resonance by a piezo-based tuning mechanism driven with a triangle-wave voltage source.

# Supplementary Information

## 1. Numerical convergence criteria

Numerical convergence criteria are developed to identify novel stable solutions under steady-state conditions. The optical intensity is analyzed to distinguish between trivial continuous-wave solutions, noise states, and the nontrivial solutions of interest. The number of round trips needed for convergence varies from less than 300 to greater than 8000. The cavity is numerically simulated for 2000 round trips, which is enough to reliably distinguish between modes of operation without excessive computation time. The numerical results are identified as continuous-wave solutions first if the difference between the minimum and maximum intensity of the waveform is <0.1 W during the last simulated round trips. Distinguishing between noisy and nontrivial converged solutions is more involved. For this analysis we examine the maximum value of the intensity during the last 100 round trips simulated. Two types of convergence are found: convergence to a constant maximum intensity value and convergence to an intensity value that varies over a constant integer (>1) number of round trips. The constant convergence case can be identified by solutions with minimal variation of the maximum intensity over the last 100 round trips and the periodically converging solutions are identified through a Fourier analysis of the intensity over the last 100 round trips. After the solutions are analyzed for convergence, the continuous-wave solutions and noisy non-converged solutions are identified by white space, and the converged solutions are identified by a color associated with a pulse parameter of interest (e.g. number of peaks in Fig. 1b and other metrics in Supplementary Information, Section 3). For a given set of system parameters, multiple solutions are possible, which can result in multiple types of convergence. Therefore, the type of convergence is a function of the simulated initial conditions.

## 2. Numerical dependence on initial conditions

The steady-state solutions of the driven-cavity system are strongly sensitive to the initial waveform used to seed the numerical simulations. Moreover, it is challenging to find the appropriate initial conditions for obtaining non-trivial steady-state solutions. Even when a stable non-trivial solution exists with a particular set of system parameters, many initial waveforms result in the trivial continuous-wave solution, which is often also stable. For example, Fig. S1b represents all the converged non-trivial solutions obtained from a single Gaussian initial waveform with a width of 3 ps and 10-W peak power. The colored region is much sparser because many of the parameters where non-trivial solutions were previously identified (Fig. S1a) now converge to the trivial continuous-wave solution instead. In this case the chirped pulses are not observed at all. If instead, a completely random white-noise initial condition is used (where each point in time corresponds to a power that varies from 0 W to a maximum power of 143 W), a different subset of nontrivial solutions is revealed (Fig. S1c), including several of the chirped pulses. Notice that neither initial conditions result in all of the non-trivial solutions. In these examples, the chirped pulses only appear with the noisy initial condition and the solutions near 2.5 rad only appear with the Gaussian initial condition. To evaluate the dependence on initial conditions in more detail, for a single set of system parameters, we examine the converged solutions for many different Gaussian initial fields parametrized by temporal width and peak power (Fig. S1d). We find a complex arrangement of solutions as a function of these initial conditions. In Fig. S1d, the white regions indicate the continuous-wave solution and the color represents the number of nontrivial pulses in the converged waveform. There is considerable variation in the character of the converged solutions for small changes in the initial condition. Therefore, to maximize the probability of obtaining nontrivial solutions, for each choice of system parameters we examine many initial conditions and retain only the nontrivial solutions. Specifically, we find a good optimization of solutions obtained vs. computation time with 15 initial conditions, including 9 random initial fields (where the power varies from 0 to 143 W) and 6 Gaussian initial fields (with 10-W peak power and a pulse width that varies in equally spaced increments from 500 fs to 3 ps). This set of initial conditions was used to obtain the solutions illustrated in Fig. S1a, for example. Owing to this sensitive dependence on initial conditions, it is likely that the results presented are incomplete and that with alternative initial conditions stable solutions could be obtained for additional parameters.

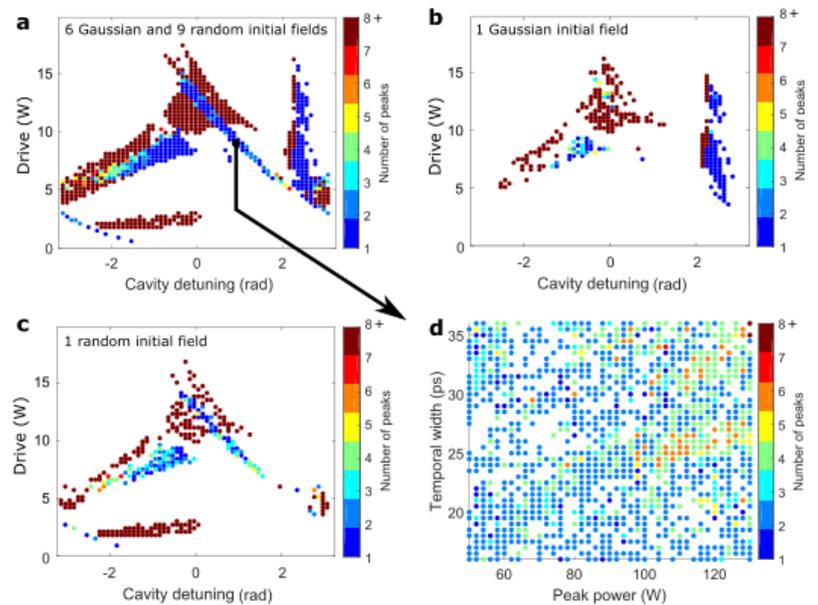

Fig. S1. **Numerical sensitivity to initial conditions. a** Converged solutions from Fig. 1b including the results from 15 different initial fields, with 9 random profiles and 6 different Gaussian profiles. The minimum number of peaks among the converged solutions is plotted for each pair of drive power and detuning values. **b** Converged solutions given a single Gaussian initial filed with 10-W peak power and 3-ps pulse width. **c** Converged solutions given a single initial condition with random intensity variations. **d** Illustration of different converged solutions as a function of the duration and peak power of the input Gaussian field for the drive frequency and detuning indicated from **a**. White space represents continuous-wave or non-converged solutions.

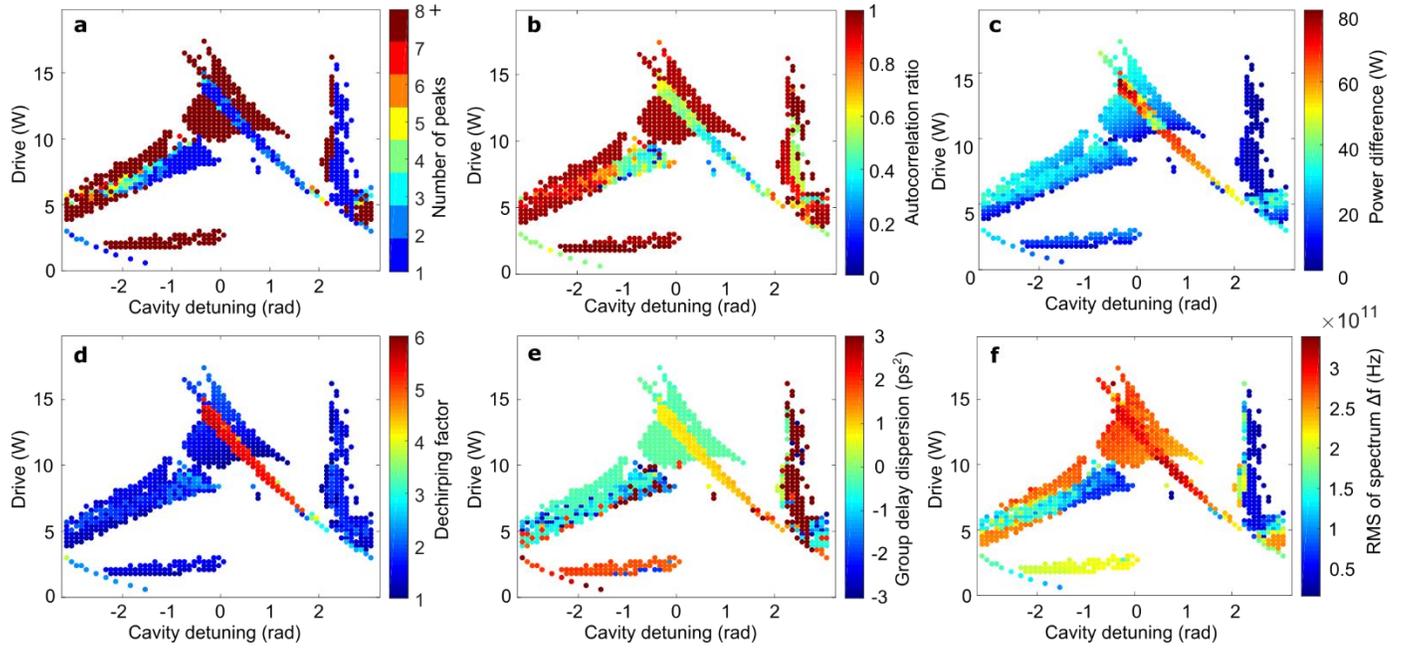

Fig. S2. **Evaluating different parameters of the steady-state solutions.** Converged solutions as a function of drive power and detuning for an all-normal dispersion cavity. The color maps represent **a** the number of prominent intensity peaks, **b** the intensity periodicity defined by the ratio of the central peak to the neighboring peak of the autocorrelation of the intensity profile (1 corresponds to a highly periodic pattern), **c** the difference between the maximum and minimum of the waveform power in the time domain, **d** the ratio of the peak power after external compression to that before, **e** the group delay dispersion required to maximize the ratio in **d**, and **f** the root-mean-squared bandwidth of the output spectrum.

## 3. Evaluating and distinguishing stable numerical solutions

The results of the numerical simulations are illustrated as a function of two experimentally convenient variables, the drive and the detuning. In Fig. S2a (and Fig. 1 of the paper), the number of intensity peaks is indicated by a color for each converged solution as a function of drive (y-axis) and detuning (x-axis). In this case, the minimum number of peaks observed from among the converged solutions seeded by 15 different initial conditions (see the previous Section) is represented in the plot. The number of intensity peaks is a useful solution parameter because it resolves different classes of solutions in parameter-space particularly well. However, other waveform parameters provide valuable and complementary information. Here we examine multiple parameters of the complex stable solutions observed in normal dispersion driven cavities with spectral filtering. The periodicity of the waveforms can be defined by the contrast ratio between the central intensity and neighboring points of a numerical autocorrelation of the waveform in time. If this ratio is one, the solution is periodic and if this ratio is zero, the solution is aperiodic. In this case, the lowest ratio from among the converged solutions seeded by 15 different initial conditions is represented in Fig. S2b. The periodicity highlights periodic Turing waves (red in Fig. S2b) as well as single pulses, such as the chirped-pulses (light-green region in Fig. S2b). While related, this information is distinct from the total number of prominent peaks (compare Fig. S2a and Fig. S2b). Because the chirped-pulse, dark soliton, and switching wave solutions each have a small number of peaks the contrast between these solutions and the Turing patterns is slightly larger for this parameter (e.g. dark blue regions of Fig. S2a). The chirped-pulse peak power is large even before dechirping, which is evident when examining the difference between the waveform maximum and minimum powers (Fig. S2c). To clearly distinguish the chirped pulses from other solutions we examine the ratio of the waveform peak power after dechirping to its peak power before dechirping and represent in Fig. S2d the maximum of this ratio from among the converged solutions obtained from all 15 initial conditions. The waveforms are dechirped by applying anomalous group-delay dispersion until the peak power is maximized. The corresponding magnitude of this dispersion is also plotted in Fig. S2e. However, while this is useful information for a waveform with a dechirping factor >2, as it is for chirped pulses, it is less useful for waveforms that do not change in peak intensity with the application of dispersion, because this implies that the pulses do not have significant or well-defined quadratic spectral phase. Finally, the spectral bandwidth (the maximum rms bandwidth from among all the converged solutions obtained from 15 initial conditions is plotted in Fig. S2f) is of interest for applications. Notably, the solutions with the largest spectral bandwidth include the chirped pulses (red) and the Turning patterns (orange). Overall, the chirped pulses are characterized by few peaks, low periodicity, a high-power difference, large dechirping factor that corresponds to 1 ps² of GDD, and broad bandwidth. Chirped temporal solitons can be distinguished through most of these parameters, as illustrated in Fig. S2.

## 4. Coexisting nonlinear solutions

Driven fiber optical cavities are complex nonlinear systems that can support a variety of stable structures. In cavities with anomalous dispersion, two commonly observed solutions are solitons and Turning patterns. Notably, these distinct nonlinear solutions have also been found to coexist, with a single soliton stabilized on a stable Turing pattern [48]. This type of nonlinear coexistence is also found in several variations in the normal dispersion system with a spectral filter examined here. In the normal dispersion system, Turing patterns are stable over large regions of parameter space (Fig. S3 red), with a particularly large region of stable solutions between 10 and 15 W of drive power. The chirped pulses, as discussed previously, are stable over a large region between 5 and 15 W of drive power (Fig. S3 blue line). Noticeably, these two regions overlap between 10 and 15 W. In the overlapping region the two nonlinear solutions coexist, and the chirped pulses exist on top of a periodic background. The chirped pulses therefore exhibit a continuous-wave background for drives smaller than 10W and an oscillating background for greater powers (Fig. S3). In addition to the Turning and chirped-pulse coexistence we also find coexistence between many other combinations of the observed stable nonlinear waveforms. One immediate consequence of this complicated coexistence behavior is the resultant difficulty in interpreting the solution parameters introduced in Supplementary Section 3.

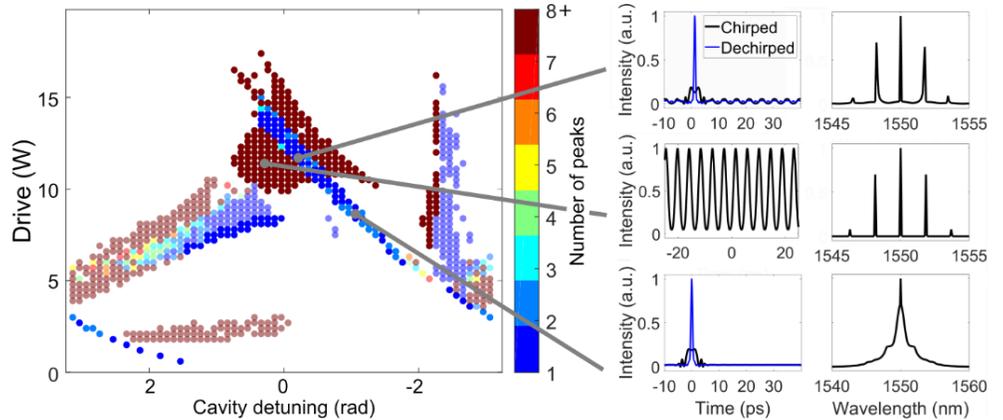

Fig. S3 **Coexistence of nonlinear solutions.** The chirped pulses (blue region and bottom right) and the Turing patterns (red regions and middle right) coexist in parameter space resulting in a nonlinear solution combining the two solutions (top right).

## 5. Dependence on the filter bandwidth

The filter bandwidth must be chosen appropriately to stabilize chirped pulses in the cavity. To evaluate the dependence of the regions of existence on the spectral filter bandwidth, numerical simulations are performed with the same parameters as the all-normal dispersion cavity (52.5 m length, Gaussian filter), but with varying filter bandwidths. Changes to the stable chirped pulses over a range of drive and detuning values can be identified with the dechirping factor (Fig. S4). Chirped pulses are not observed without a spectral filter in the cavity. Chirped pulses begin to appear for Gaussian filters with full-width at half maximum bandwidths of 8 nm or narrower. Full width at half maximum bandwidths between 6 and 4 nm enable chirped pulses with high dechirping factors over a broad range of detuning values. This range therefore defines the optimum filter bandwidths for this cavity. The threshold for chirped pulses is approximately 5 W with a 6-nm filter bandwidth. With smaller drive powers, switching waves are observed instead. The threshold decreases with narrower spectral filter bandwidths. For example, the threshold is 2.5W with a 2-nm filter, where the switching waves obtained with the broader filter become chirped pulses. Narrower filter bandwidths reduce the threshold further but with a corresponding decrease in the bandwidth and the chirped-pulse compression ratio. From Eq. 1, the stability regions obtained for a given filter (e.g. 6 nm) can be recovered with a different filter bandwidth (e.g. 8 nm) by making a corresponding change in the total group delay dispersion (e.g. with $(6/8)^2$ times less dispersion).

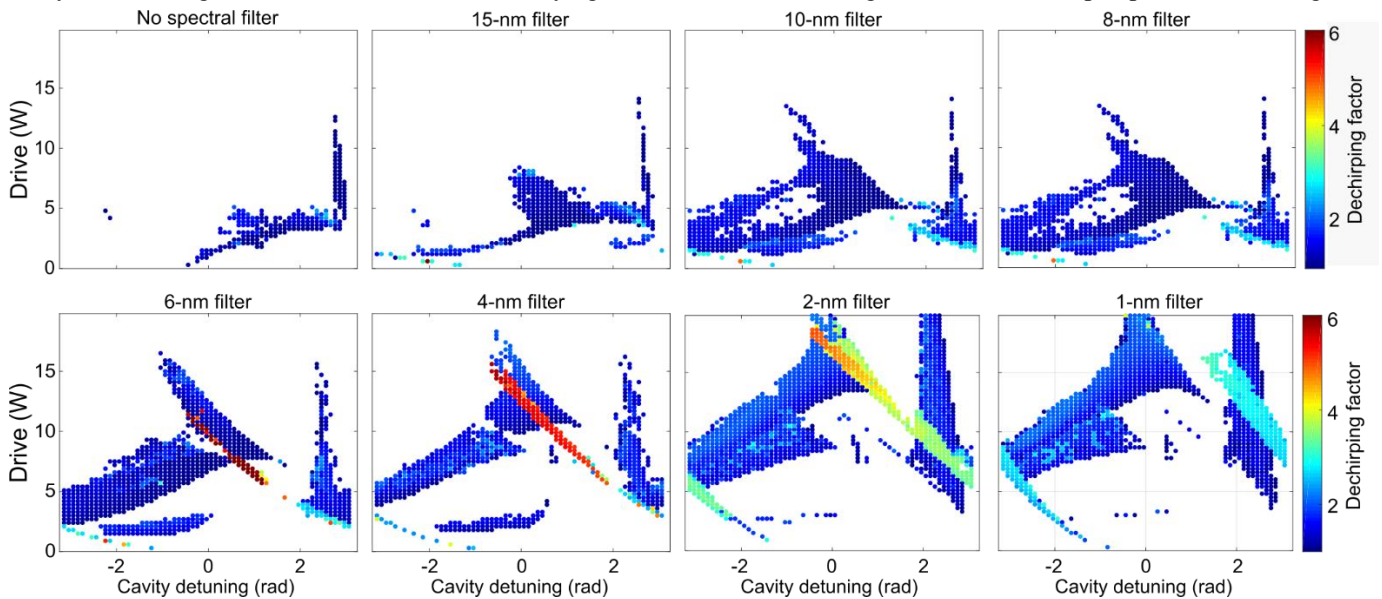

Fig. S4. **Chirped temporal soliton existence vs. spectral filter bandwidth.** The dechirping factor for converged solutions vs. drive power and detuning for different bandwidths, $\Delta\lambda$ (full width at half the maximum), of a Gaussian spectral filter for a cavity length of 52.5 m. Stable chirped pulses appear for filter bandwidths less than 8 nm. Higher compression factors are observed with broader filter bandwidths (red).

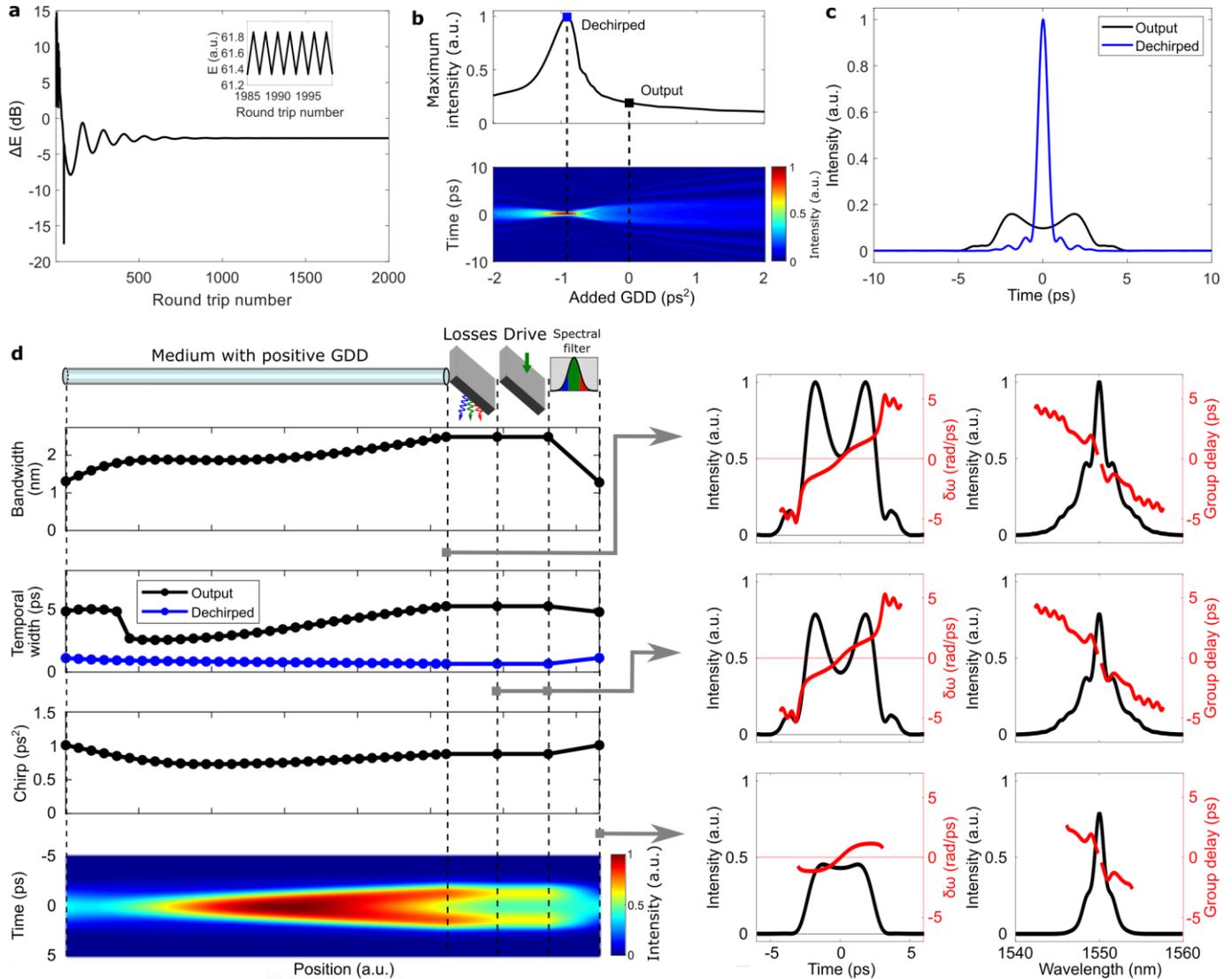

Fig. S5. **Characteristics of chirped temporal solitons with a broader, 4-nm filter. a** Numerical convergence of the pulse energy difference between subsequent round trips, $\Delta E$, to a stable numerically-limited steady-state value. The pulse energy of the last 15 round trips is plotted inset to depict the period-2 oscillation. **b** The change in the pulse and peak intensity as a function of group-delay dispersion applied after the cavity, indicating compression with anomalous dispersion with a maximum at GDD = -0.9 ps$^2$. **c** The chirped cavity output (black) and dechirped (blue) pulses from b. **d** Evolution of steady-state chirped temporal soliton bandwidth (FWHM), temporal width (FWHM), chirp (defined by the GDD required to maximize the pulse intensity, with the opposite sign), and pulse intensity in the cavity. The FWHM of the pulse after dechirping the pulse at each position of the cavity is plotted in blue. The associated pulse intensity, instantaneous frequency, power spectrum, and group delay from the indicated locations in the cavity are plotted on the right.

## 6. Chirped cavity soliton dynamics

The dynamics of the chirped temporal solitons depend on the parameters of the cavity. Fig. 2 of the paper depicts the evolution of chirped solitons from an all-normal dispersion cavity with a 2-nm Gaussian spectral filter, a drive of 11.4 W, and a detuning of 1.36 rad. This configuration yields one of the least complex evolution dynamics among the chirped-pulses solutions observed. In this section, we describe examples of more complex dynamics observed for chirped-pulse solutions with different parameters. While the essential characteristics are the same including the consistently large positive chirp and broad spectral bandwidth, the details of the evolution such as the pulse compression factor and the solution periodicity, can vary.

*Broader filter bandwidth.* The filter bandwidth is a critical parameter in determining the character of the solutions. For comparison with the results of a 2-nm spectral filter (Fig. 2), we examine a solution with a 4-nm spectral filter (Fig. S5). This solution exists with 9.6-W drive power and 0.62-rad detuning. An interesting difference with this solution can be seen in the convergence (Fig. S5a). The solution converges to a periodic solution alternating between slightly different peak powers (<1% variation) every other round trip. While qualitatively distinct, this small change has a negligible effect on the pulse dynamics. Interestingly, for this filter bandwidth only solutions with this 2-period convergence are observed (after 2000 round trips). The chirp is similar to the previous case which corresponds to a GDD of approximately 0.9 ps$^2$ with a dechirping factor of ~6 (Fig. S5b-c). The broader bandwidth associated with the broader filter results in a shorter compressed pulse with a duration of 600 fs. The qualitative evolution of the spectrum is the same, with a bandwidth increase in the waveguide due to nonlinear broadening and a decrease from the spectral filter (Fig. S5d). The temporal pulse duration evolution is also qualitatively similar with an overall increase due to dispersion but with a steeper drop of the pulse duration in the first fiber section. Overall the chirped pulses observed with different filter bandwidths are qualitatively similar but with small quantitative differences. In this case the broader filter bandwidth supports a chirped temporal soliton with a broader bandwidth.

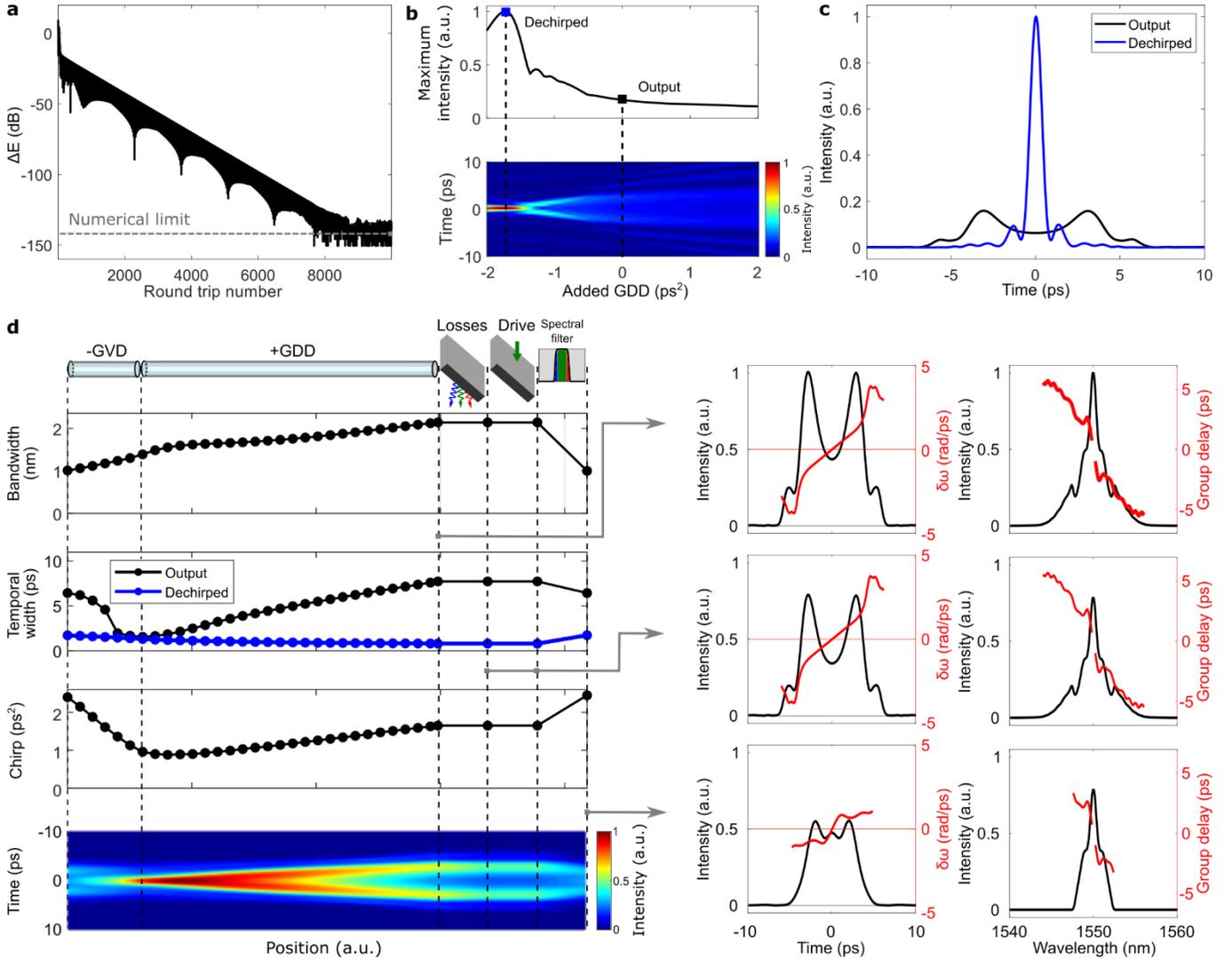

Fig. S6. **Characteristics of chirped temporal solitons from Region i in the net-normal dispersion cavity. a** Numerical convergence of the pulse energy difference between subsequent round trips, $\Delta E$, to a stable numerically-limited steady-state value. **b** The change in the pulse and peak intensity as a function of group-delay dispersion applied after the cavity, indicating compression with anomalous dispersion with a maximum at GDD = -1.7 ps$^2$. **c** The chirped cavity output (black) and dechirped (blue) pulses from b. **d** Evolution of steady-state chirped temporal soliton bandwidth (FWHM), temporal width (FWHM), chirp (defined by the GDD required to maximize the pulse intensity, with the opposite sign), and pulse intensity in the cavity. The FWHM of the pulse after dechirping the pulse at each position of the cavity is plotted in blue. The associated pulse intensity, instantaneous frequency, power spectrum, and group delay from the indicated locations in the cavity are plotted on the right.

*Chirped pulses in a net-normal dispersion cavity.* As described in the paper, a longer cavity is beneficial for reducing the drive threshold and increasing the peak power of the pulsed drive source. A 150-m cavity was designed with a net-normal dispersion equivalent to that of the shorter 52.5-m cavity. This was achieved through introducing anomalous dispersion fiber   Additionally, motivated by the experimental parameters the filter is implemented as a 12$^{th}$ order super-Gaussian spectral filter with a FWHM bandwidth of 4.25 nm and the round-trip loss is 1.05 dB. The solutions from this cavity are represented in Fig. 1d for comparison with the all-normal dispersion case (Fig. 1b) in the main paper. The additional fiber sections may also impact the evolution of the chirped pulses in the cavity. We examine these differences for the chirped pulses from Regions i and ii from Fig. 1d.

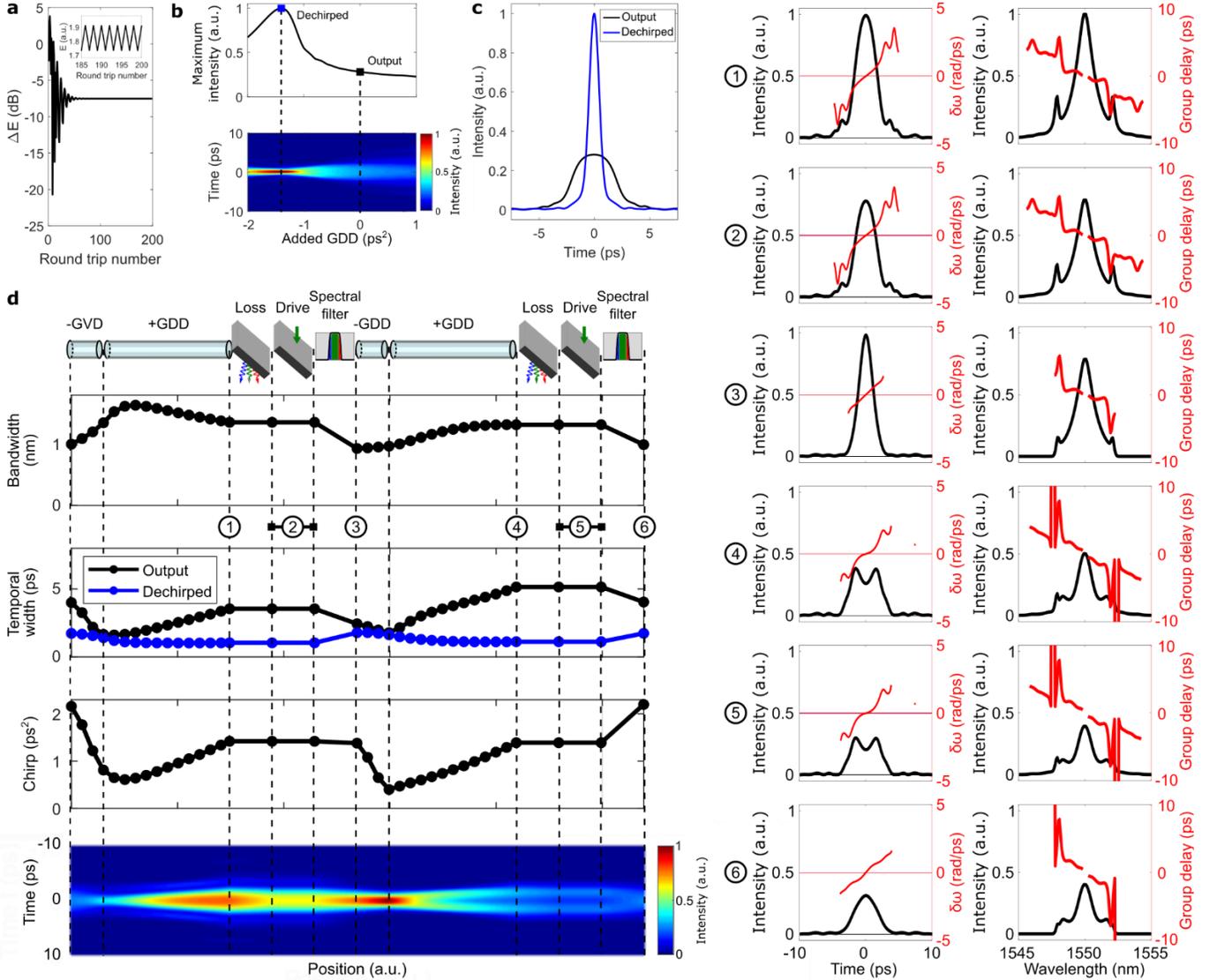

Fig. S7. **Characteristics of chirped temporal solitons from Region ii in the net-normal dispersion cavity. a** Numerical convergence of the pulse energy difference between subsequent round trips, $\Delta E$, to a stable numerically-limited steady-state value. The pulse energy of the last 15 round trips is plotted inset to depict the period-2 oscillation. **b** The change in the pulse and peak intensity as a function of group-delay dispersion applied after the cavity, indicating compression with anomalous dispersion with a maximum at GDD = -1.4 ps$^2$. **c** The chirped cavity output (black) and dechirped (blue) pulses from b. **d** Evolution of steady-state chirped temporal soliton bandwidth (full width at half the maximum, FWHM), temporal width (FWHM), chirp (defined by the GDD required to maximize the pulse intensity, with the opposite sign), and pulse intensity in the two last round trips of the cavity. The FWHM of the pulse after dechirping the pulse at each position of the cavity is plotted in blue. The associated pulse intensity, instantaneous frequency, power spectrum, and group delay from the indicated locations in the cavity are plotted on the right.

*Region i.* Fig. S6 depicts pulses with a drive of 2.95 W and a detuning of 0.9 rad in chirped-pulse Region i. This solution requires 8000 round trips to reach numerically-limited convergence (Fig. S6a) and the dechirping factor remains similar, with a value of ~6 (Fig. S6b-c). The spectral bandwidth increases in both of the waveguide sections through nonlinear broadening and then decreases with the application of the filter. The pulse duration now decreases in the anomalous dispersion fiber as the pulse is partially dechirped (Fig. S6d). The residual pulse broadening is compensated by a reduction through spectral filtering. The chirp is larger than in the all-normal dispersion cavity, and corresponds to a GDD of 1.7 ps$^2$ on the pulse output before the spectral filter. This is in part because of the additional normal dispersion fiber that is required to form the additional dispersion map. The pulse (with two peaks) and the spectrum look qualitatively similar to those in the all-normal dispersion case. Overall, this solution is very similar to that from the all-normal dispersion cavity with small changes originating from the additional segments of fiber.

*Region ii.* Fig. S7 depicts pulses with a drive of 3.8 W and a detuning of 0.43 rad in chirped-pulse Region ii. In this case the solution requires two round trips before the evolution repeats. Unlike in Fig. S5 where the pulse varies by <1% between round trips, in this case the variation is >10% (Fig. S7a). In Fig. S7b-c and Fig. 3 of the paper, the average of the dechirped pulse from each round trip is plotted. The averaged dechirping factor is slightly smaller than with a peak power increase of ~4 after dechirping. The large variation of peak power every round trip requires evaluating the pulse evolution over two subsequent round trips in Fig. S7d. While subtle differences can be identified, the overall evolution remains qualitatively the same: the spectral bandwidth and pulse duration have a net increase in the fiber sections and are both decreased back to their initial values after spectral filtering. The chirp from pulses in Region ii is large throughout the evolution, with an output chirp of 1.4 ps$^2$. Experimentally, the average between the output waveforms from each of the two round trips is measured. While the simulated chirped-pulse solutions are all qualitatively similar, the solutions from Region ii have the closest quantitative agreement with the chirped pulses observed experimentally.

## 7. Chirped-pulse scaling laws

Simple scaling laws relating solution parameters to the system parameters can be derived from a master equation model of the cavity. The damped and detuned driven nonlinear Schrödinger equation, or the Lugiato-Lefever equation (LLE), is an established model for the driven nonlinear optical cavity without a filter. The spectral filter essential for chirped pulse generation is modeled with an additional term with a second derivative with respect to time, which corresponds to a Gaussian spectral filter. The LLE with a spectral filter for a slowly varying electric field envelope A is given as

$$L\frac{\partial A}{\partial z} = -\alpha A + i\delta A + \left(i\frac{\overline{\beta_2}}{2}L + \frac{1}{f^2}\right)\frac{\partial A}{\partial t^2} + i\gamma L|A|^2 A + \sqrt{D}, \tag{1}$$

where z is the propagation coordinate, t is the fast time, L is the length of the cavity, $\alpha$ corresponds to the cavity loss, $\delta$ is the round-trip detuning, $\overline{\beta_2}$ is the average group-velocity dispersion, $\gamma$ is the Kerr nonlinear parameter, $f$ is the spectral filter bandwidth, and $D$ is the intracavity drive power. By normalizing z, t, and A in the following way,

$$z' = \frac{z}{z_0} = \frac{z}{L}, \quad \tau = \frac{t}{T_0} = t\sqrt{\frac{\alpha}{L|\overline{\beta_2}|}}, \quad \text{and} \quad u = A\sqrt{\frac{1}{P_0}} = A\sqrt{\frac{\gamma L}{\alpha}}, \tag{2}$$

Eq. 1 can be rewritten as

$$\frac{1}{\alpha}\frac{\partial u}{\partial z'} = -u + i\frac{\delta}{\alpha}u + \left(i\frac{\text{sgn}(\overline{\beta_2})}{2} + \frac{1}{L|\overline{\beta_2}|f^2}\right)\frac{\partial^2 u}{\partial \tau^2} + i|u|^2 u + \sqrt{\frac{D\gamma L}{\alpha^3}}. \tag{3}$$

By equating the left-hand side to zero to account for steady-state conditions, Eq. 3 becomes

$$0 = -u + i\frac{\delta}{\alpha}u + \left(i\frac{\text{sgn}(\overline{\beta_2})}{2} + \frac{1}{L|\overline{\beta_2}|f^2}\right)\frac{\partial^2 u}{\partial \tau^2} + i|u|^2 u + \sqrt{\frac{D\gamma L}{\alpha^3}}. \tag{4}$$

The resultant normalized equation,

$$0 = -u + i\delta_0 u + \left(\frac{i}{2} + \frac{1}{f_0^2}\right)\frac{\partial^2 u}{\partial \tau^2} + i|u|^2 u + \sqrt{D_0}, \tag{5}$$

is defined in terms of three unitless parameters corresponding to the drive ($D_0$), spectral filter ($f_0$), and detuning ($\delta_0$), given by

$$D_0 = D\frac{\gamma L}{\alpha^3}, \quad f_0 = f\sqrt{L|\overline{\beta_2}|}, \quad \text{and} \quad \delta_0 = \frac{\delta}{\alpha}. \tag{6}$$

If chirped temporal solitons are known to be stable in a cavity with specific values of $D_0$, $f_0$, and $\delta_0$, stable chirped solitons can be obtained for a different cavity if the values for the unitless coefficients do not change. The first relationship from Eq. 6 reveals that the required drive power has a cubic dependence on the total cavity loss and an inverse linear dependence on the total cavity nonlinearity (see Supplementary Information, Section 8). The second relationship conveys that the filter bandwidth must scale inversely with the square root of the total group delay dispersion (see Supplementary Information, Section 5). The third relationship suggests that the same solution can be recovered if the relative drive frequency scales linearly with the cavity loss.

The peak power, pulse duration, and energy also scale according to the system parameters, as defined by Eq. 2:

$$T_0 = \sqrt{\frac{L|\bar{\beta}_2|}{\alpha}}, \qquad P_0 = \frac{\alpha}{\gamma L}, \quad \text{and} \quad E = T_0 P_0 = \frac{\alpha}{\gamma L}\sqrt{\frac{L|\bar{\beta}_2|}{\alpha}} = \frac{1}{\gamma}\sqrt{\frac{\alpha|\bar{\beta}_2|}{L}}. \tag{7}$$

Eq. 7 reveals that broader bandwidths can be achieved with smaller dispersion and that large pulse energies are achieved with smaller nonlinearity and larger dispersion. The scaling laws from Eq. 6 enable efficient design of cavities that can support experimentally realizable chirped temporal solitons. They can be readily applied to very different parameter regimes and platforms, including for micro-scale comb generation on-chip, or for macroscopic enhancement cavities.

## 8. Dependence on the cavity length

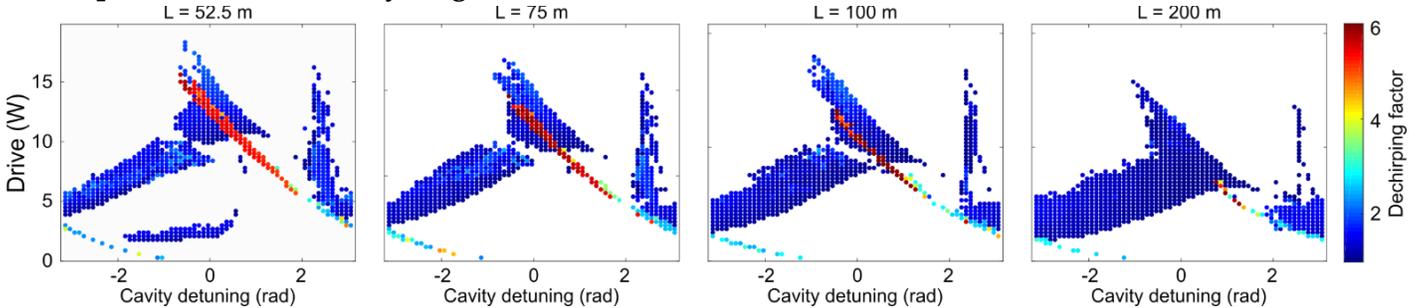

Fig. S8. **Chirped temporal soliton existence vs. cavity length.** Illustration of the dechirping factor for converged solutions vs drive power and detuning for different cavity lengths, L, with a constant 4-nm spectral filter. Increasing the cavity length has a similar effect to increasing the spectral filter bandwidth: stable chirped pulses become sparser in parameter-space.

The length of the cavity determines the total nonlinearity and the group-delay dispersion. From Eq. 6, changes to the total nonlinearity change the drive threshold and changes to the group-delay dispersion change the required filter bandwidth. To examine and verify these effects, we examine the dependence of the regions of existence on the cavity length with numerical simulations, with all other parameters held constant (Fig. S8). The upper drive limit varies from 15 W to 10 W for a cavity that is 50% longer, in agreement with the predicated linear relationship. As the length of the cavity is increased further, the number of parameter values where chirped pulse solutions exist decreases dramatically. This can be understood from the relation between the filter bandwidth and the cavity length in Eq. 6. From Fig. S4 and Supplementary Information Section 5, we found that chirped pulses are not observed when the filter is too large. In other words, when the normalized filter bandwidth, $f_0$, is too large, stable solutions do not exist. From Eq. 6, increasing the cavity length also increases $f_0$ and so we would also not expect to find stable chirped pulse when the cavity is too long. However, this relation also reveals that for long cavities, a value of $f_0$ that supports stable chirped solitons can be recovered with a spectral filter with a narrower bandwidth. Simulations of the dependence of the solutions on length and on spectral filter bandwidth confirm the validity Eq. 6.

## 9. Driving the cavity with pulses

High drive powers are required to observe stable chirped temporal solitons. The required high drive powers can be achieved by modulating the drive into nanosecond pulses before amplification. The effective drive power is then enhanced by the ratio of the drive pulse duration to the cavity period, which enable two orders of magnitude enhancement for the parameters used in this study. To confirm that the drive enhancement scales as expected, pulse pumping is applied to a cavity which produces traditional solitons in the anomalous dispersion regime (Fig. S9a). The continuous-wave drive power threshold for a 107-m long anomalous dispersion cavity is approximately 0.25 W of average power. By decreasing the duty cycle of the drive, the amplified peak power increases which should enable a corresponding decrease in the average power required by the amplifier. The required average power requirement decreases as expected (Fig. S9b), which validates the pulsed drive approach experimentally.

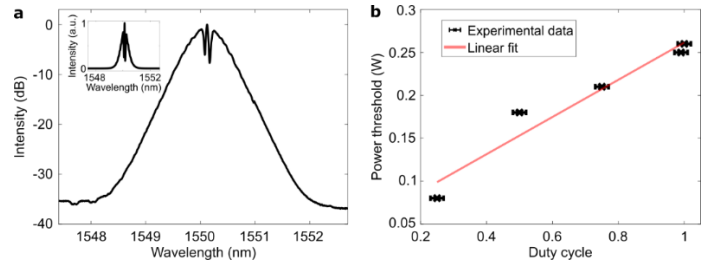

Fig. S9. **A pulsed drive source enables higher drive powers**. **a** Output spectrum on a log scale of experimentally observed anomalous dispersion solitons, with the linear scale inset. **b** Minimum average power of the amplifier required for observing stable solitons vs. the duty cycle of the drive, illustrating a linear increase of the drive peak power.

## 10. Simulation and measurements of the nonlinear cavity resonance

The cavity resonance contains a large amount of global information about the complex nonlinear system. In the paper, a single resonance from experiment is examined and compared to an equivalent numerical simulation, with good qualitative agreement. The goal of this section is to provide more information about this complex resonance, including additional experimental data and corresponding numerical results. Experimentally, the resonance is obtained from the time-integrated cavity output intensity as a function of the drive frequency when swept through the cavity resonance. The data from Fig. 3 corresponds to a drive average power of 1.8 W (incident on the cavity, which corresponds to 68 W of peak drive power coupled into the cavity). Here we examine the resonance behavior for drive average powers that vary from 100 mW to 2 W (Fig. S10). At each power we adjust the polarization of the drive to align to a single polarization state of the cavity by minimizing additional resonances. The output of the cavity is observed by detecting it on a photodiode after filtering out any spectral content outside of the central 20-GHz drive band with a fiber Bragg grating filter. The peak of the experimental resonance is shifted to match the peak from the equivalent numerical sweep. The numerical sweeps are obtained with noise-seeded simulations in which the detuning is varied after every round trip at a rate determined by the experimental sweep time.

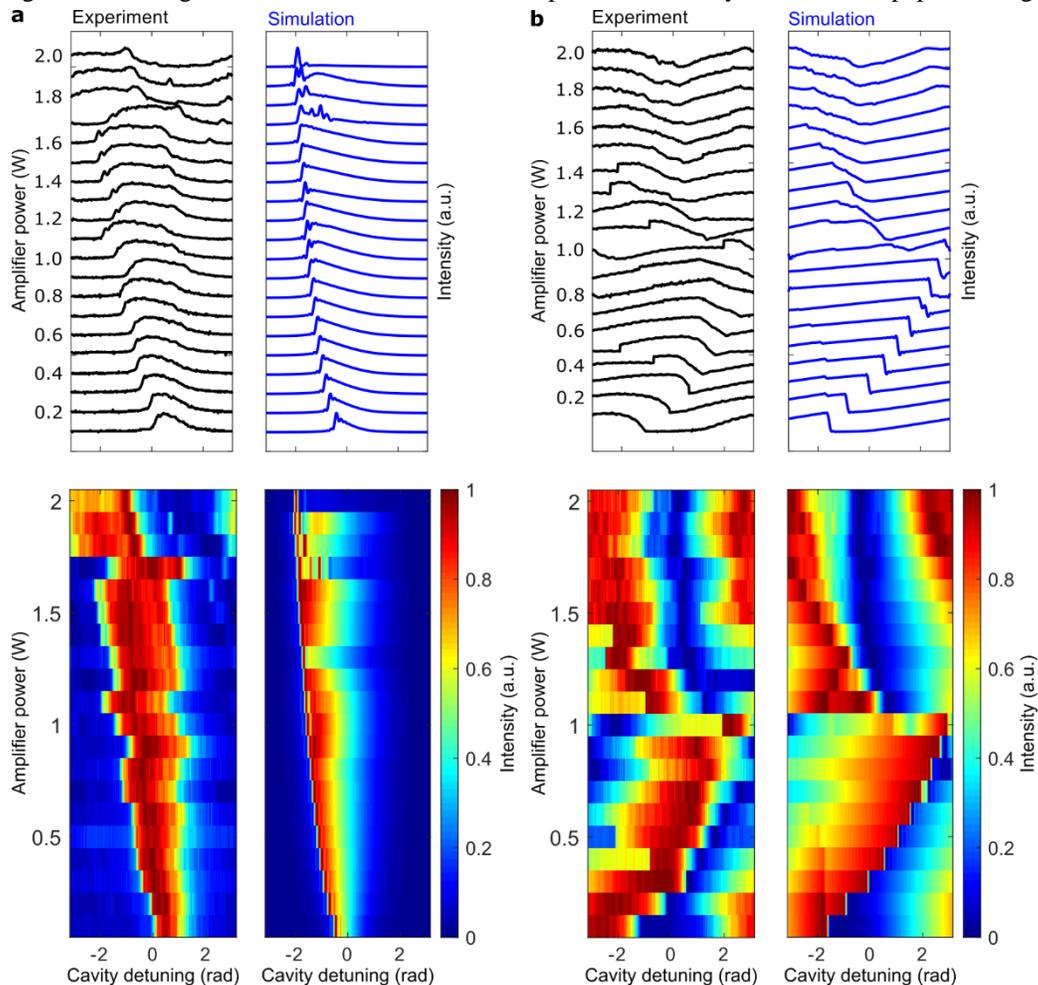

Fig. S10. **Cavity resonance measurements and corresponding simulations.** (Top) Experimental (black) and simulated (blue) intracavity continuous wave power as a function of the drive frequency detuning and average power **a** from negative to positive and **b** from positive to negative detuning. (Bottom) Color map representation of the same results. The resonance contains global information about the complex nonlinear system. Hysteresis is observed as a function of sweep direction.

Given the triangle wave used to drive the experimental frequency sweep, the resonance is swept from positive to negative detunings as well as from negative to positive detunings. With increasing detunings (Fig. S10a), the resonance peaks become broader with higher drive. They are also shifted to negative detuning values. With decreasing detunings (Fig. S10b), the resonance behaves differently and is no longer around zero detuning at low power. The resonance is also broader and exhibits more structure. The qualitative agreement with experiment is in general good. There are also features that do not agree due to the complexity of this system. For example, there is some disagreement at higher drives in Fig. S10a. Also, the step-like structures are inconsistent at around 1.1-W drive in Fig. S10b. There are several plausible explanations for these discrepancies related to environmental perturbations, polarization dynamics, pulse pumping dynamics, and the sensitive dependence on initial conditions. Since the cavity resonances are not actively stabilized, environmental thermal and acoustic perturbations will distort the resonance observed. Any coupling between orthogonal polarization states in the cavity will produce additional resonances within this sweep range. In addition, dynamics resulting from the other polarization state can strongly affect the overall nonlinear dynamics and consequently the shape of the resonance. The present simulation neglects the polarization dynamics and assumes a single polarization state. Experimentally, the cavity is driven by a 10-ns drive pulse.

In the numerical simulations a continuous-wave drive is assumed because the drive pulse is much longer than the simulated temporal window. However, even a slight change in the drive magnitude as a function of time can lead to noticeable effects in the nonlinear dynamics of the system which would translate to changes in the resonance features. Finally, as emphasized in Supplementary Information Sections 2 and 4, the numerical simulations are complicated and feature multiple stable states, small basins of attraction, and a sensitive dependence on initial conditions. These complexities are all expected to be encountered when continuously traversing all of the different values of detuning. Many different classes of solutions, including dark solitons, switching waves, Turing patterns and chirped pulses can exist simultaneously within the same parameter space. In addition, each of these solutions can be obtained with different initial conditions, whereas in the simulation of the sweep, only one initial condition is chosen. We apply random intensity waveforms for the starting initial condition and by varying the detuning seed the next solution with the waveform that is obtained from the previous detuning. However, in practice, any small perturbations in the system will lead to different nonlinear pathways. In other words, sweeping over the parameters of a system sensitive to initial conditions further enhances this sensitivity, which complicates the analysis. Overall, it is remarkable to obtain qualitative agreement for resonances obtained from this highly nonlinear and complex system. Closer agreement may be obtained with a more complicated model that addresses the challenges summarized here.

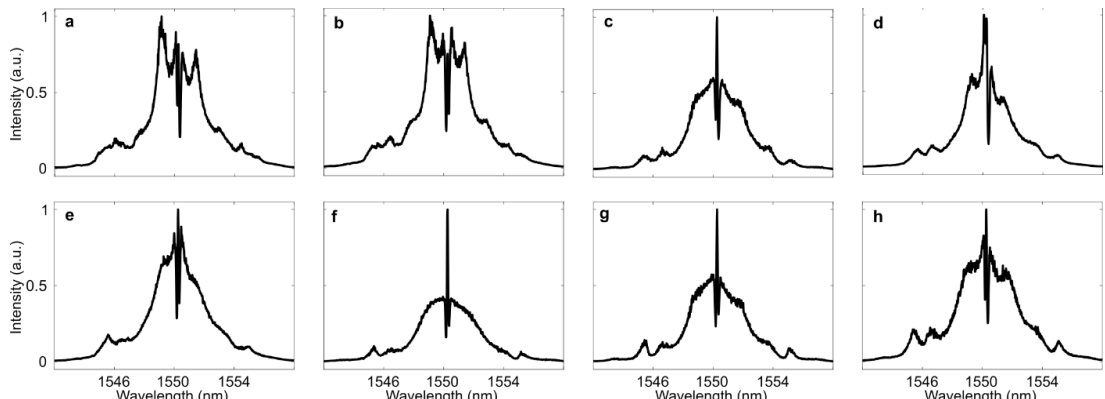

Fig. S11. **Variation of the observed chirped-pulse output spectra.** Example variations in the output spectra include **a-b** steeper sides, **c-d,g-h** smooth sides with four side-peaks and **e-f** smooth sides with two side peaks.

## 11. Variation in the chirped-pulse output spectra

A range of chirped-pulse output optical spectra are observed experimentally (Fig. S11). Small changes in the polarization state, power, and frequency of the drive result in subtle changes to the chirp-pulse spectrum. Several of the observed spectra are plotted in Fig. S11 to give a more complete representation of the experimental data than the single spectrum from Fig. 3 in the paper. Small changes observed include variations in the ratio of the central drive wavelength to the spectrum of the pulse, variation in the prominence of the side peaks, and variation of the concavity of the central part of the spectrum. All of the observed spectra qualitatively agree with the numerically simulated spectra. Additionally, a comprehensive analysis of the numerical simulations, including a vectorized system of equations accounting for both polarization states of the fiber, could help account for some of these variations.

## 12. Analysis of the intensity autocorrelations

For a collinear two-photon intensity autocorrelation the ratio of the detected signal peak to the background is dependent on the pulses as well as the residual continuous-wave background. Without the continuous-wave background, this peak-to-background autocorrelation ratio is 3 to 1 (this is the case for the pulses output from mode-locked lasers, for example). However, as the continuous-wave background increases, this ratio reduces. We illustrate this effect numerically with a Gaussian pulse on a continuous-wave background with a variable ratio of the relative amplitude of these two contributions (Fig. S12). We find that the autocorrelation contrast noticeably decreases when the amplitude of background to pulse peak approaches 0.5. In contrast, the ratio increases if there are more pulses in the cavity because the ratio of pulses to total background increases. It also increases when the chirped pulses compress because the pulse peak power increases. The autocorrelation background therefore conveys useful information about the chirp and pulse compression. In addition to the pulse duration (Fig. 3f) the autocorrelation background is also minimized when the pulse is dechirped (Fig. S12d). For comparison with numerical simulations, several assumptions are needed. The numerical temporal window is sampled over 100 ps to minimize computation time. Therefore, the background is extrapolated up to the full 10-ns duration of the drive pulse. After leaving the cavity, the background is filtered with a fiber-Bragg grating and then amplified with an Erbium-doped fiber amplifier before autocorrelation measurements. We find agreement with the experimental results with a net 15-dB attenuation of the background after these elements (Fig. S12e). In summary, the autocorrelation ratio contains additional information regarding the complex waveform output from driven resonators.

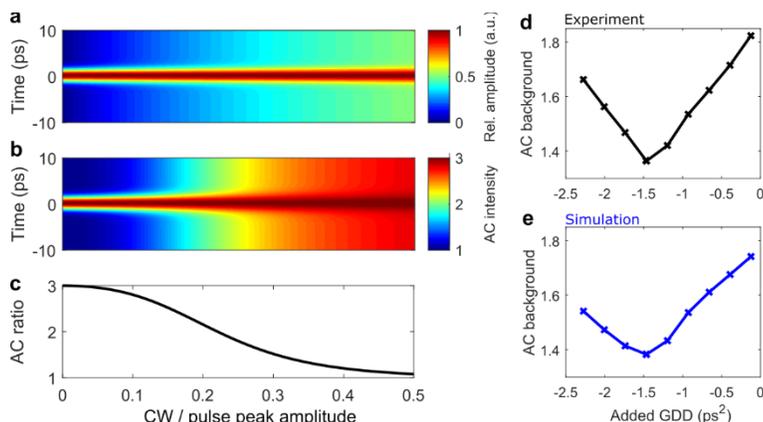

Fig. S12. **Simulated autocorrelation of a pulse with a continuous background and comparison with experiment. a** Pulse intensity profile, **b** autocorrelation profile and **c** autocorrelation ratio with the ratio of continuous-wave amplitude to the peak pulse amplitude of the waveform varying along the horizontal axis. **d** Experimental and **e** simulated autocorrelation ratio of the chirped temporal solitons as a function of the GDD from the grating pair compressor with a minimum corresponding to the fully dechirped pulse.

## 13. Relationship between chirped temporal solitons and other normal dispersion waveforms

In normal dispersion resonators, researchers have examined dark pulses[33–35], bright pulses[36], switching waves[32], platicons[37,38], and travelling front solutions[32]. Moreover, these solutions have been shown to be closely related to each other[32,37]. Here we explore the relationship between the chirped pulse solutions and previous solutions in the normal dispersion regime. Optical fibers are a nearly ideal waveguide for studying pulse propagation and so non-ideal effects, such as nonlocal coupling and higher-order group velocity dispersion are neglected. We examine and reproduce a subset of the solutions from Ref. [30] based on the Lugiato Lefever equation (LLE), to facilitate the comparison between different solutions. Numerically simulated solutions of the LLE (without a filter) are plotted in parameter space with respect to the analytical solutions for the continuous-wave solutions. The continuous-wave solutions are given by

$$D_0^{\pm}(\delta_0) = \tfrac{1}{3}\left(2\delta_0 \pm \sqrt{\delta_0^2 - 3}\right)\left[1 + \tfrac{1}{9}\left(\sqrt{\delta_0^2 - 3} \pm \delta_0^2\right)\right], \tag{8}$$

where $D_0^{\pm}$ is the normalized drive and $\delta_0$ is the normalized detuning. These expressions for the drive power are plotted as a function of detuning in green in Fig. S13. For reference, the critical value of the equilibria of the electric field squared, $\beta = D_0(\delta_0) = 1 + (1 - \delta_0)^2$, is plotted as a dashed line[30]. We additionally reproduce the dark pulse solutions from Ref. [30] over a well-defined line in the detuning vs. drive parameter space. At low drive powers, dark solitons are observed. At higher drive powers the solutions develop a more complex structure. At higher drive powers the solutions become unstable and the location of this boundary agrees well with the results in Ref. [30]. By introducing a spectral filter, even one with a broad 20-nm bandwidth, the location of the solutions within the parameter space shifts and the character of the solutions begin to change. The more complex dark solitons become stable and can exist with ten times higher drive powers (Fig. S13b). Moreover, the distinction between dark solitons, platicons, and switching waves begins to become less clear. In comparison to chirped solitons, these solutions do not have a quadratic spectral phase (or chirp) and their 3-dB bandwidth is narrow. If the bandwidth of the spectral filter is narrow enough (e.g. 4-6 nm as described in Supplementary Information Section 5) and the drive is sufficient, the bandwidth increases, the pulse duration decreases, and pulses with a well-defined chirp can form (Fig. S13b). Given the close relation in parameter space and the continuous evolution from dark solitons to chirped pulses, it is clear that the chirped pulses are related to the other solutions in normal dispersion resonators. Pulse formation in the normal dispersion cavities has been described as the result of interlocking switching-waves or front solutions [32]. Interestingly, the chirped pulse solutions in mode-locked lasers can also be described as the interlocking of traveling front solutions. Chirped-pulse mode-locked lasers are well described by the cubic quintic Ginzburg Landau equation which is known to possess chirped pulse solitons described in this way [46]. While this relationship is suggestive, a thorough theoretical analysis will be needed to fully describe chirped temporal solitons in passive cavities. Moreover, the complete set of solutions and dynamics for normal dispersion resonators with a filter is highly nontrivial (e.g. see Fig. 1 and S2). For example, in addition to the branch of solutions examined in Fig. S13, dark solitons, switching waves, and other solutions are also stable in completely different parameter regimes (Fig. 1 and S2). A comprehensive theoretical investigation for all of these nontrivial solutions will be important for future devices based on this novel filtered resonator design.

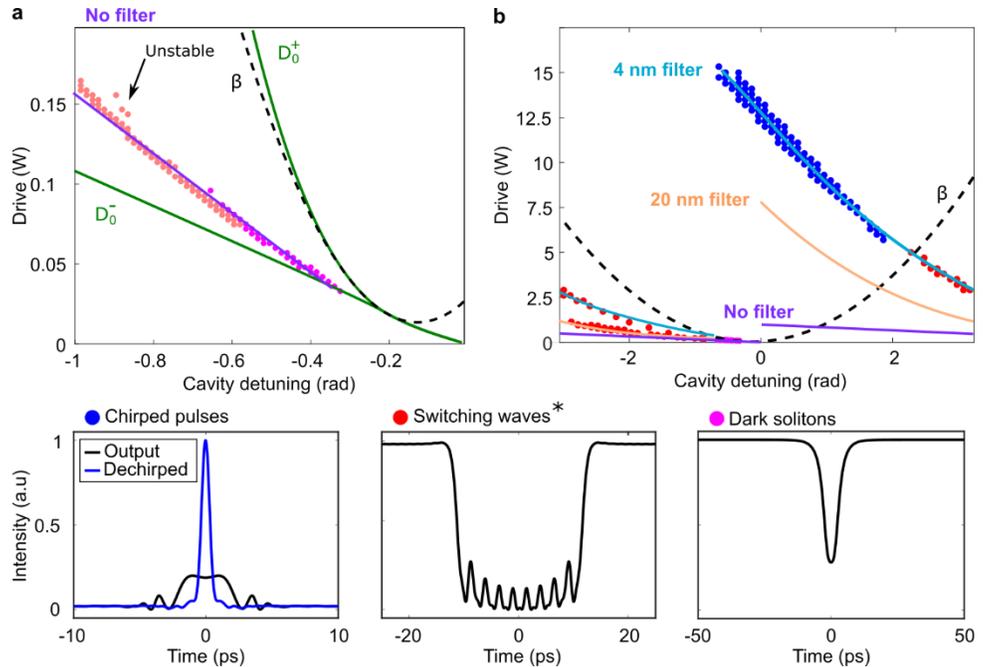

Fig. S13. **Relation in parameter space between solutions in normal dispersion cavities. a** Dark solitons in a normal-dispersion cavity without a spectral filter just above the analytically calculated lower branch $D_0^-$ in excellent agreement with Ref. [30]. For drive powers > 0.1 W, solitons become unstable and do not converge. **b** With a 20-nm Gaussian spectral filter, dark solitons are found at low drive powers and complex dark solitons at higher drive powers. For powers >0.5 W, dark solitons appear as switching-waves. Switching waves evolve to chirped pulses featuring a decreasing peak duration and an increasing spectral bandwidth with narrower spectral filtering (light blue). Solid lines represent phenomenological fits to guide the reader. *These solutions have been referred to as switching waves, interlocking front solutions, or long complex dark pulses.

## 14. Comparison of chirped solitons to solitons in anomalous-dispersion cavities

In mode-locked lasers, chirped solitons stabilize high pulse energies. In general, when the pulse is chirped, its peak power remains low, which reduces the destabilizing effects of nonlinearity. In normal dispersion resonators, chirped pulse mode-locked lasers have enabled pulse energies that are as much as two orders of magnitude larger than what can be achieved with traditional solitons [27–29]. Numerical simulations can help determine the relative energy of chirped-pulse solitons in passive resonators. We begin with the chirped-soliton resonator parameters for a 52.5-m normal dispersion fiber and a 4-nm bandwidth spectral filter. To examine comparable traditional solitons, we change the sign of the dispersion and remove the spectral filter. We find stable solitons over a well-defined region of drive powers and detuning values. As the drive power increases, the continuous-wave-background, and consequently the solitons, begin to destabilize. We select the soliton that is noise-free and stable with the largest drive power as the high-performance representative for traditional solitons. The energy of the resultant pulse corresponds to 15 pJ inside the resonator with a drive power of 0.3 W and a detuning of -1.34 radians (Fig. S14a). In the normal dispersion cavity, chirped pulses exist over a wide range of drive powers and detuning values. However, the chirped pulses have a drive power threshold below which chirped pulses are not stable. For the 4-nm filter, chirped pulses are not stable at the 0.3-W drive power of the noise-free, stable traditional soliton. However, with a narrower 1-nm spectral filter, it is possible to stabilize chirped pulses with only 0.3-W of drive power. The resultant chirped pulses have pulse energies of 100 pJ, or about seven times more energy than the solitons (Fig. S14b). This result suggests that chirped pulses also carry a significant energy benefit in Kerr resonator systems.

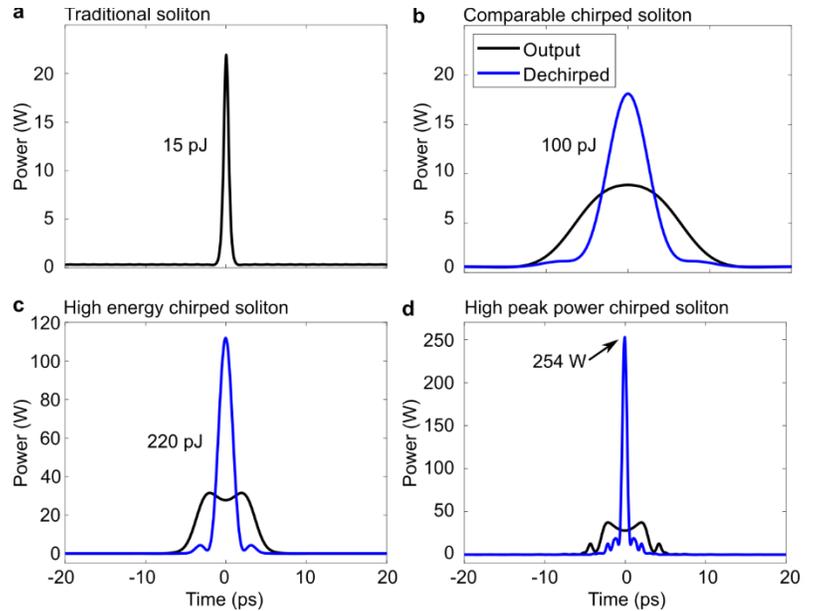

Fig. S14. **Chirped temporal soliton energy and peak power in comparison to traditional solitons.** Stable soliton solution from numerical simulations of an **a** all-anomalous dispersion cavity without a filter and an **b** all-normal dispersion cavity with a 1-nm filter. The chirped-pulse soliton has around seven times higher pulse energy than the traditional soliton with the same drive power and magnitude of dispersion. **c** At higher drive powers, chirped temporal solitons can support another two times higher energy than the traditional soliton. **d** High peak powers can be achieved after dechirping chirped temporal solitons obtained with a spectral filter with a larger 6-nm bandwidth.

In the previous result, the chirped-pulse drive power was constrained for direct comparison between the two types of solitons. However, higher energies may be possible for the chirped-pulses with higher drive powers. To investigate, simulations are run for all possible drive detuning values as well as for much larger drive powers. Stable solutions are found for powers as much as fifty times higher than for the comparable anomalous dispersion cavity. The chirped-pulse energy is found to increase with increasing drive. The bandwidth also increases with the drive power. Clean, noise-free pulses with energies of at least 220 pJ are observed in this cavity (Fig. S14c). This corresponds to more than ten times the energy of traditional solitons. However, the bandwidth of this chirped pulse is narrower than that of the soliton. By increasing the spectral filter, the bandwidth can be increase. For example, with a 6-nm spectral filter, we observe stable chirped pulses with a peak power enhancement that is greater than ten compared to traditional solitons (Fig. S14d). Numerical simulations therefore strongly suggest that much higher pulse energies and peak powers are achievable with chirped solitons than with the traditional solitons. Higher energy pulses for frequency combs corresponds to a higher power per comb line, which is an important performance parameter.